%% file: Main.tex
\documentclass[12pt]{article}
\usepackage{amsmath}
\usepackage{graphicx}
\usepackage{enumerate}
\usepackage{url}

\usepackage[a4paper, lmargin=2.7cm, rmargin=2.5cm, bottom=2.5cm, top=2.3cm]{geometry}
\usepackage{amsmath, amsthm, amssymb}
\usepackage{graphicx}
\graphicspath{{Figures/}}
\usepackage[round]{natbib}
\usepackage{float}
\usepackage{placeins}

\usepackage{graphicx}
\graphicspath{{./Images/}{./Figures/}}

\usepackage{subcaption}
\newtheorem{theorem}{Theorem}[section]

\newtheorem{remark}[theorem]{Remark}

\theoremstyle{definition}

\newtheorem{definition}[theorem]{Definition}

\usepackage{hyperref}

\usepackage{xr}
\usepackage{./macros/mymacros}
\usepackage{./macros/packagemacros}

\usepackage{xcolor}

\title{Robust FWER control in Neuroimaging using Random Field Theory: Riding the SuRF to Continuous Land Part 2}
\author{Samuel Davenport, Armin Schwartzman, Thomas E. Nichols\\ and Fabian J.E. Telschow}
\begin{document}
	
\maketitle

\begin{abstract}
	\input{./abstract.tex}

\end{abstract}

\section{Introduction}
\input{./introduction.tex}

\input{./main_body.tex}

\section{Discussion}\label{SS:discussion}
\input{./discussion.tex}
\section*{Acknowledgements}
\input{./acknowledgements.tex}

\bibliographystyle{plainnat}
\bibliography{./bibfiles/RFT,./bibfiles/MHT,./bibfiles/Statistics,./bibfiles/TomsPapers,./bibfiles/fMRI,./bibfiles/Books}
\input{./appendix.tex}

\end{document}

%% file: abstract.tex
Historically, applications of RFT in fMRI have relied on assumptions of smoothness, stationarity and Gaussianity. The first two assumptions have been addressed in Part 1 of this article series.
Here we address the severe non-Gaussianity of (real) fMRI data to greatly improve the
performance of voxelwise RFT in  fMRI group analysis. In particular, we introduce a
transformation which accelerates the convergence of the Central Limit Theorem allowing
us to rely on limiting Gaussianity of the test-statistic. We shall show that, when the
GKF is combined with the Gaussianization transformation, we are able to accurately
estimate the EEC of the excursion set of the transformed test-statistic even when
the data is non-Gaussian. This allows us to drop the key assumptions
of RFT inference and enables us to provide a fast approach which correctly controls
the voxelwise false positive rate in fMRI. We employ a big data \cite{Eklund2016}
style validation in which we process resting state data from 7000 subjects from
the UK BioBank with fake task designs. We resample from this data to create realistic
noise and use this to demonstrate that the error rate is correctly controlled.\\
\textit{Keywords:} Random Field Theory, FWER control, multiple testing, voxelwise inference, spatial statistics, non-stationarity, Gaussianization.

%% file: introduction.tex
Random Field Theory (RFT) encompasses an advanced set of mathematical techniques for analysing imaging data that has been widely applied in neuroimaging in order to control false positives via voxel, cluster and peak level inference (\cite{Worsley1992}, \cite{Worsley1996}, \cite{Friston1994}, \cite{Chumbley2009})\footnote{In voxelwise inference voxels with test-statistic values lying above a multiple testing threshold are determined to be significant. In both cluster and peak level inference a cluster defining threshold is used to identify clusters/peaks and then thresholding based on cluster extent and peak magnitude is used to determine significant clusters/peaks.}. RFT has traditionally required that the data is Gaussian, stationary and sufficiently smooth \citep{Worsley1996}, however none of these assumptions is particularly reasonable in fMRI. \cite{Eklund2016} demonstrated that RFT inference can fail to control false positive rates as a result of the failure of these assumptions. It is thus highly desirable to extend RFT to work without these requirements. There has been considerable work on extending RFT to work under non-stationarity (see \cite{Taylor2006}, \cite{Adler2007}, \cite{TelschowHPE}, \citep{Telschow2023FWER}, \cite{Cheng2015}), but it has never been implemented in neuroimaging toolboxes or shown to work using realistic validations. Here we provide an accurate and fast voxelwise RFT framework that does not require stationarity nor smoothness and no longer relies on the data being Gaussian.

For this article we shall call the body of work of Worsley, Friston, Taylor and Adler on RFT inference as \textbf{Traditional RFT}. This refers to the framework that was built in \cite{Worsley1992}, \cite{Friston1994}, \cite{Worsley1996}, \cite{Taylor2006} and \cite{Taylor2007}. See \cite{Ostwald2021} for an overview of the application of RFT in fMRI. Traditional RFT inference makes a key informal assumption, colloquially referred to as the {\em good lattice assumption}, that observations on the lattice corresponding to the brain have the same properties as a continuous random field \citep{Worsley1996}. This is required in order to provide good estimates of the smoothness of the data \citep{Kiebel1999} and to correctly infer on the distribution of the maximum of the voxelwise test-statistic field. Failure of this assumption leads to voxelwise inference being conservative \citep{Hayasaka2003}. In order to adequately satisfy the good lattice assumption, traditional RFT requires a high level of smoothing, much higher than the levels typically applied in fMRI \citep{Worsley2005}. Generally data is smoothed with an isotropic Gaussian kernel with FWHM between $ 2 $ and $ 4$  voxels \citep{Woo2014} however in fMRI, as we shall show, traditional RFT inference is conservative even when the applied FWHM is $ 6 $ voxels. We addressed these issues with the RFT framework in Part 1 of this article series by introducing the
concept of Super Resolution Fields (SuRFs) \cite{Telschow2023FWER}. After applying a smoothing kernel,
the data used in fMRI can be viewed as a \textit{convolution field} which is a special case of a SuRF.

Traditional RFT requires that the data is a Gaussian random field or that the voxelwise test-statistic is a Gaussian (related) random field. Given a large enough number of subjects the later is guaranteed to approximately hold by the \textit{Central Limit Theorem} (CLT). However the sample sizes typically used in fMRI are not large enough to make this a reasonable assumption. Using 7000 subjects from the UK biobank we will show that fMRI data can have heavy tails meaning that the CLT takes longer to converge and that in practice a large number of subjects is required before the test-statistic is approximately Gaussian. The lack of Gaussianity in fMRI data has been observed by others in the literature. In particular \cite{Wager2005}, \cite{Woolrich2008} and \cite{Fritsch2015} considered robust regression models in order to account for non-Gaussianity and \cite{Chen2012}, \cite{Roche2007} developed methods for dealing with outliers. More recently \cite{Parlak2023} observed ``clear violations of the assumption of Gaussianity" in first level data. In the RFT literature, \cite{Worsley2005} and \cite{Schwartzman2019} used transformations which transform the test-statistic to Gaussianity but still required that the underlying fields be Gaussian. \cite{Nardi2008} and \cite{TelschowSCB} showed that RFT provides asymptotically valid inference so long as the test-statistic is asymptotically Gaussian, but rely on the CLT holding to provide valid inference, see also \cite{Telschow2020Delta}. So far it has not been shown that the type of non-Gaussianity that can be present in fMRI can be made compatible with RFT, especially at the sample sizes typically used in practice. To address this we propose an approach which transforms the data to improve levels of Gaussianity, under the null hypothesis, via a process called Gaussianization. We show that using this, in combination with the RFT inference framework, improves the validity of RFT theory by making it robust to non-Gaussianity.

RFT has historically required stationarity (\cite{Adler1981}, \cite{Worsley1992}, \cite{Worsley1994}, \cite{Worsley1996}). This allows for estimation of the quantiles of the maximum of a random field, via the \textit{expected Euler characteristic} (EEC) \citep{Worsley1992}, and is used to control the \textit{familywise error rate} (FWER). Under non-stationarity, \cite{Taylor2006} established the \textit{Gaussian Kinematic Formula} (GKF) for the EEC which only depends on the \textit{Lipshtiz Killing Curvatures (LKCs)} - constants which depend solely on the covariance function of the data. This approach was incorporated into the RFT inference framework in \cite{Taylor2007b} who used a warping estimate to calculate the LKCs. More recently \cite{Adler2017} and \cite{TelschowHPE} have used regression and projection techniques to estimate the LKCs. However these non-stationary approaches have never been implemented in commonly used neuroimaging toolboxes (e.g. SPM, FSL, AFNI). In \cite{Telschow2023FWER} we developed a SuRF estimator of the LKCs which, as we shall show, can be applied to neuroimaging data to calculate the LKCs. We shall show that, when combined with the Gaussianization transformation and convolution fields, we can use this estimator to accurately estimate the EEC of the excursion set of the transformed test-statistic.

\cite{Eklund2016} pre-processed resting state data using fake task designs in order to determine the validity of the traditional RFT inference framework. This type of resting state validation creates realistic noise on which to test the validity of methods for performing inference in fMRI. In what follows we shall rather to this procedure as the \textit{Eklund test}, a terminology introduced in \cite{Lohmann2018}. Historically, prior to the work of \cite{Eklund2016}, RFT has primarily been validated using stationary Gaussian simulations (\cite{Hayasaka2003}, see also \cite{Davenport2022Ravi}). As \cite{Eklund2016} showed, these types of simulations do not adequately test the robustness of RFT inference and are not sufficient to establish validity of these types of inference methods in practice. The Eklund test represents a gold standard in terms of method validation and has been used in other works, e.g. \cite{Lohmann2018} and \cite{Andreella2023}, to establish the false positive rates of their procedures. So far it has been applied by sampling subjects from datasets consisting of at most 198 subjects which causes problems that we address here by using a significantly larger dataset.

 

The primary contributions of this work are several fold. Firstly we have pre-processed resting state data from 7000 subjects from the UK biobank in order to perform realistic validation of our methods. Secondly, we show that a large of the conservativeness of traditional RFT inference in fMRI is due to the evaluations of the field on the lattice and can be fixed by using convolution fields as proposed in Part 1. Thirdly we demonstrate that, even when convolution fields are used, the GKF provides a biased estimate for the expected Euler characteristic, causing substantial additional conservativeness. We show that the driving cause of this is the heavy tailed nature of the noise. Fourthly, in order to address this, we introduce a transformation which corrects for non-Gaussianity and accelerates the convergence of the CLT. When applied to the transformed data we show that the GKF can be used to closely estimate the upper tails of the EEC curve. Finally we validate our methods using the Eklund test based on the 7000 subjects to show that the EEC is well estimated and that the FWER is correctly controlled.



Software to implement the methods (including estimation of LKCs, calculating coverage rates, performing Gaussianization and other RFT analysis) is available in the RFTtoolbox package \citep{RFTtoolbox}. Code to reproduce the analyses and figures of the paper is available at \url{https://github.com/sjdavenport/ConvolutionFieldsNeuro}.
	

%% file: main_body.tex
\section{Theory}
In this section we discuss how the RFT inference framework from Part 1 can be adapted in order for it to be applied in neuroimaging. The resulting modifications shall allow for valid inference, even when the data is non-Gaussian. 

In what follows we will let $ \mathcal{V} \subset \mathbb{R}^D$, for some $ D \in \mathbb{N} $, be a regular lattice consisting of a finite number of points which corresponds to the set of voxels that make up the brain\footnote{In brain imaging we typically have $ D = 3 $, however, the theory is fully general so we state it for arbitrary $ D $. For instance applications on cortical surface data have $ D = 2 $.}. Here we shall use the word \textit{voxel} to refer to the elements of $ \mathcal{V} $ (treating voxels here as points in $ \mathbb{R}^D $ rather than boxes). We will assume that $ \mathcal{V} $ is equally spaced in the $ d $th direction with spacing $ h_d $ for $ d = 1, \dots, D. $ We consider group level analyses, in which we have a number of $ N > 2$ such that for $ n = 1, \dots, N $ the $ n $th subject has a corresponding 3D random image $ X_n = \mu_X + \epsilon_n $ where $ \mu_X: \mathcal{V} \rightarrow \mathbb{R} $ and $ (\epsilon_n)_{n = 1}^N $ are i.i.d. symmetric random fields on $ \mathcal{V} $ with bounded variance\footnote{In neuroimaging $ X_n $ is the $ c^T\hat{\beta} $ COPE (contrast of parameter estimates) image obtained from running pre-processing and first level analyses for each subject \citep{Mumford2009}. }. For each $ v \in \mathcal{V} $, define the sample mean $ \hat{\mu}_N(v) = \frac{1}{N}\sum_{n = 1}^N X_n(v) $ and sample variance $ \hat{\sigma}_N^2(v) = \frac{1}{N-1}\sum_{n = 1}^N \left( X_n(v) - \hat{\mu}_N(v) \right)^2. $


\subsection{Improving robustness to non-Gaussianity}\label{SS:Gauss}
RFT inference relies on Gaussianity (\cite{Worsley1996}, \cite{Adler2007}). As we will demonstrate in our large scale validations in Sections \ref{SS:RSV}, this assumption seems not to hold for fMRI datasets and in practice the noise distribution can be very heavy tailed. Without Gaussianity we are no longer able to 
correctly calculate the EEC using the GKF. Given large enough numbers of subjects it is generally reasonable to assume that the test-statistic field is approximately a Gaussian random field because of the CLT. Under this assumption we can still apply the GKF directly to the resulting test-statistic field \citep{Nardi2008} and so RFT based FWER control, as described in Part 1, will be asymptotically valid. However, the sample sizes used in fMRI are usually to small to rely on the CLT, as we show in Section \ref{SS:RSV}. In particular, if the voxelwise distributions are heavy tailed, as they can be in fMRI, the rate of convergence of the CLT can be rather slow. 



In order to resolve the problems caused by the non-Gaussianity of the data we propose
a Gaussianization procedure that transforms the data so that it becomes more Gaussian. This procedure estimates a null distribution from the data, using information from across all voxels.
This distribution is then used to transform the original data in order to eliminate heavy tails and improve the level of marginal Gaussianity at the voxels where the null hypothesis is true. If we knew the marginal CDF of $ X_n(v) $: $ \Psi_v $ under the null at a given voxel $ v $ we could transform pointwise to obtain $ X_n'(v) = \Phi^{-1} \Psi_v(X_n(v)) $ to ensure that the unsmoothed data is marginally Gaussian under the null, reducing the convergence issues caused by heavy tails. The marginal CDF is typically unknown in fMRI, however, we can estimate it by taking it to be the same at each voxel (up to scaling by the standard deviation).

More formally, at each voxel $ v $ we standardize and demean the fields $ X_n $. This yields standardized fields:
\begin{equation}\label{eq:marginalnull}
X_n^{S,D} = \frac{X_n - \hat{\mu}_N}{\hat{\sigma}_N},
\end{equation}
where the operations are computed voxelwise. In order to calculate the marginal null distribution we combine this data over all voxels and subjects to obtain a null distribution. This null distribution is illustrated in Figure \ref{fig:marginalnull} using null contrast maps from 50 subjects from the UK BioBank (pre-processed as described in Section \ref{SS:rsprocess}). From this figure we can see that the resulting density is heavy tailed and narrow peaked, illustrating that the underlying distribution is non-Gaussian. The non-Gaussianity is further illustrated via the failure of the GKF to calculate the EEC correctly, see Figure \ref{fig:EECplotsFWHM5}.

To address this non-Gaussianity we standardize voxelwise without demeaning, then determine the quantile of every voxel relative to this null distribution and finally use this quantile to convert the data to have approximately Gaussian marginal distributions. More precisely, we define
\begin{equation}\label{eq:standard}
X_n^{S} = \frac{X_n}{\hat{\sigma}_N}
\end{equation}
and for each voxel $ v $ and subject $ n $ we compare $ X_n^{S}(v) $ to the null distribution to obtain a quantile
\begin{equation*}
q_n(v) = \frac{1}{N\left| \mathcal{V} \right|}\sum_{v' \in \mathcal{V}}\sum_{n = 1}^N 
\mathds{1}\big[X_n^{S}(v) \leq X_n^{S,D}(v')\big].
\end{equation*}
The Gaussianized fields are then given by
\begin{equation*}
X_n^G(v) = \Phi^{-1}(q_n(v))~~~ v\in \mathcal{V}, ~n =1,\ldots,N.
\end{equation*}
Here $ \Phi $ is the CDF of the normal distribution. 
\begin{remark}
	The assumption that the pointwise distribution of $ \epsilon_n $ is symmetric ensures that $ \mu_X(v) = 0 $ implies that $ \mathbb{E}(X_n^G(v)) = 0 $ and so we can use the Gaussianized data to infer on $ \mu_X.$ The Gaussianization does not however assume that the marginal distributions at each voxel are the same, in order to ensure asymptotic validity, as this is ensured by the CLT given sufficiently many subjects.
\end{remark}



\subsection{Gaussianized Convolution Fields}\label{SS:convfields}
In this section we define Gaussianized convolution fields which allow us to combine the SuRF framework introduced in Part 1 with the Gaussianization procedure introduced in Section \ref{SS:Gauss}. This will resolve the remaining conservativeness with the RFT framework caused by non-Gaussianity of the data.
\subsubsection{Defining Gaussianized convolution fields}
In fMRI the smoothed images are evaluated on the lattice $ \mathcal{V} $, however the original data and the smoothing kernel $ K: \mathbb{R}^D \rightarrow \mathbb{R} $ together contain unused information. In particular, for each $ n = 1, \dots, N $, given a bounded set $ S \subset \mathbb{R}^D $ such that $ \mathcal{V} \subset S$ we define the
\textbf{ Gaussianized convolution field} $ Y^G_n:S \rightarrow \mathbb{R} $ such that for $ s \in S $, 
\begin{equation}\label{eq:Gconvfield}
Y^G_n(s) = \sum_{v \in \mathcal{V}} K(s-v)X^G_n(v)
\end{equation}
which has pointwise mean $ \mu_G(s) = \mathbb{E}(Y^G_1(s)) $.

The data which is typically used for inference in fMRI is $ \lbrace Y_n:v \in \mathcal{V},\, 1 \leq n \leq N\rbrace $, where $ Y_n(s) = \sum_{v \in \mathcal{V}} K(s-v)X_n(v) $, $ s \in S $, are the convolution fields obtained by smoothing the untransformed data which we introduced in Part 1. Evaluating the fields on the lattice $ \mathcal{V} $ like this is the norm in current software and, as explained in Section \ref{SS:conserve} in the context of voxelwise RFT, causes the resulting inference to be conservative.

\subsubsection{Defining the domain and the added resolution}\label{SS:voxdomain}
\begin{figure}[h]
	\begin{center}
		\includegraphics[width=0.32\textwidth]{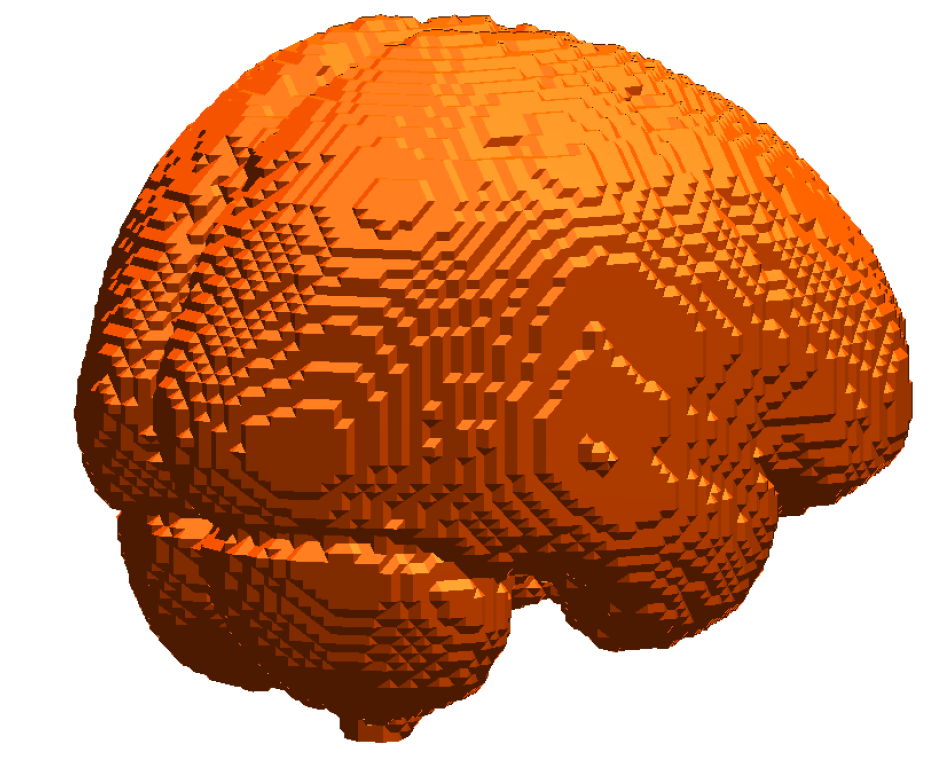}\quad
		\includegraphics[width=0.3\textwidth]{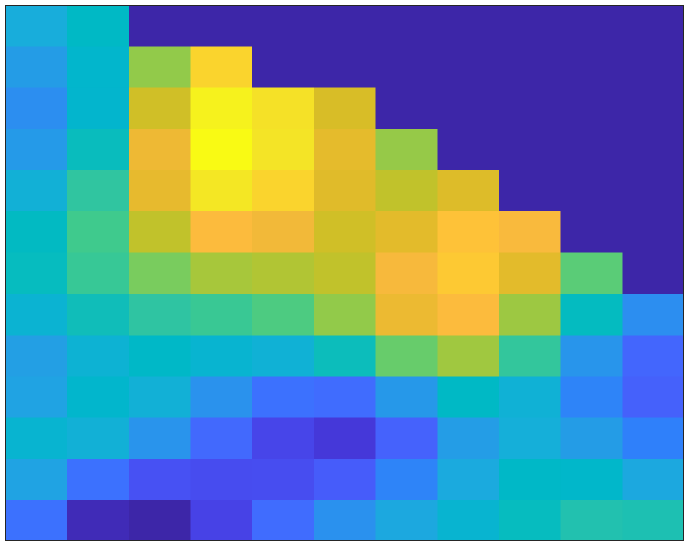}\quad
		\includegraphics[width=0.3\textwidth]{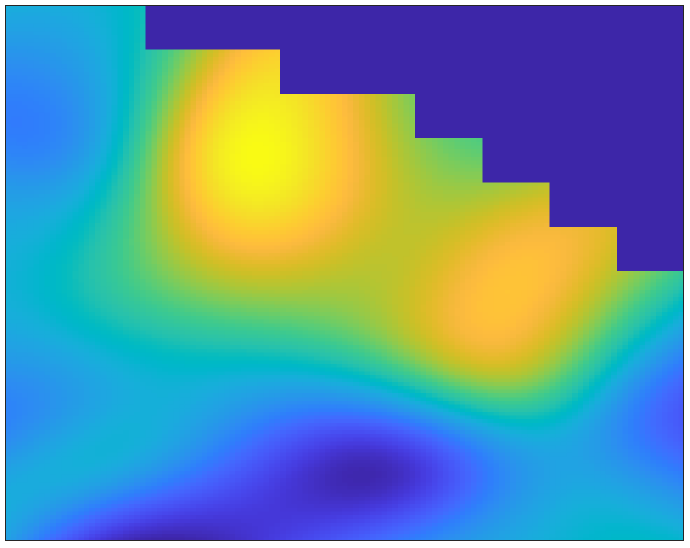}
	\end{center}
	%
	%
	\caption{Left: The voxel-domain corresponding to the standard 2mm MNI mask. It consists of the boxes that make up the mask and all the points in $ \mathbb{R}^3 $ that are within them. Middle: A 2D slice through the one-sample $ t $-statistic of 50 resting state COPE images smoothed with FWHM = 3 voxels on a lattice. These images are pre-processed as discussed in Section \ref{SS:UKB}. Right: the corresponding convolution field of the same section of the brain (evaluated on $\mathcal{V}_r$ for $ r = 11 $). The value at the centre of the voxel of the convolution field is the same as the value of that voxel on the original lattice. Points in dark blue in the upper right of the images are points that lie outside of the mask. }\label{fig:convvslatimage}
\end{figure}

As the domain $ S $ of the convolution field we use the natural choice which consists of the collection of voxels that make up the brain. I.e. for $ v = (v_1, \dots, v_D)^T \in \mathcal{V} $, let 
\begin{equation*}
S(v) = \left\lbrace (s_1, \dots, s_D)^T\in \mathbb{R}^D: \max_{d \in 1, \dots, D} \left| s_d - v_d \right| \leq \frac{h_d}{2} \right\rbrace,
\end{equation*} 
where $ h_d $ is the distance between the voxels in $ \mathcal{V} $ in the $ d $th direction. Then we define 
\begin{equation*}
S = \bigcup_{v \in \mathcal{V}} S(v).
\end{equation*}
This corresponds to a continuous version of the the mask used in fMRI, which consists of a box around each voxel $ v \in \mathcal{V} $. We call this canonical choice of $ S $ the voxel manifold and will use this as our domain throughout the remainder of the paper. An illustration of the voxel manifold for the MNI mask is shown in the left panel of Figure \ref{fig:convvslatimage}.

\begin{figure}[h]
	\begin{center}
		\includegraphics[width=0.2\textwidth]{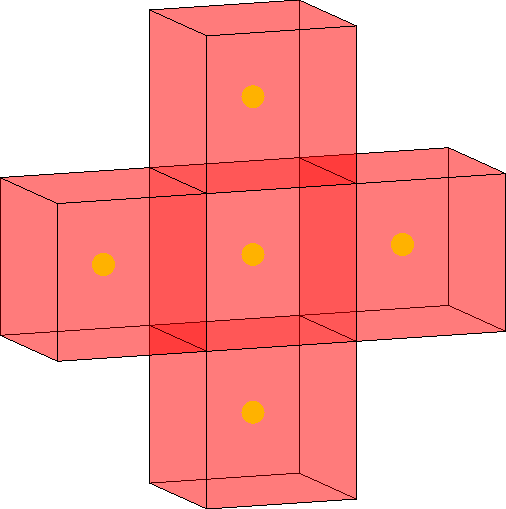}
		\quad
		\includegraphics[width=0.2\textwidth]{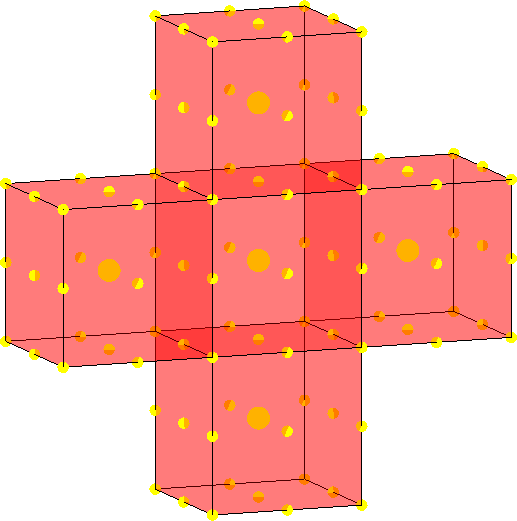}
		\quad
		\includegraphics[width=0.2\textwidth]{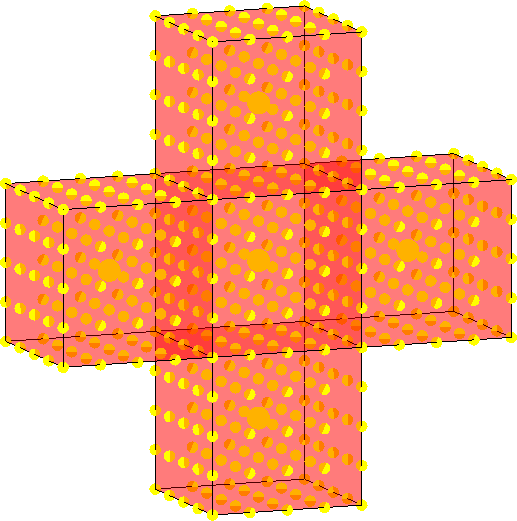}
	\end{center}
	\caption{Illustration of the added resolution for a simple 3D mask consisting of 5 voxels. The collection of red boxes corresponds to the voxel-domain for the mask. Within these, from left to right, we plot in yellow the set of points of $ \mathcal{V}$, $ \mathcal{V}_1$ and $\mathcal{V}_3 $ respectively. The large yellow dots are the locations of the original set of voxels in $ \mathcal{V}$ which are at the centres of the red boxes. The remaining yellow points indicate the points that are added.}\label{fig:resaddcomp}
\end{figure}

As in practice it is impossible to evaluate convolution fields everywhere,
we define the fine grid
\begin{equation*}
\mathcal{V}_r = \left\lbrace s \in \mathbb{R}^D: s = v + \frac{k\cdot h}{r+1} \text{ for some } k \in \mathbb{Z}^D \cap \left[ -\frac{r+1}{2}, \frac{r+1}{2} \right]^{D}\text{ and } v \in \mathcal{V}\right\rbrace
\end{equation*}
where the \textbf{added resolution} $ r \in \mathbb{N}$ is the number of points between each voxel that is introduced (taking $ r = 0 $ we obtain the original lattice, i.e. $ \mathcal{V}_0 = \mathcal{V} $). Here $ h = (h_1, \dots, h_D)^T $ is the vector of the spacings between voxels and $ \cdot $ denotes the dot product. We illustrate the difference between the $ \mathcal{V}_r $ in Figure \ref{fig:resaddcomp} for a simple 3D mask consisting of $ 5 $ voxels. Further illustrations for other 3D masks are available in Figures \ref{fig:singlevox}, \ref{fig:2by2}, and for a 2D slice through the MNI mask in Figure \ref{fig:2DMNI}.

\subsubsection{Testing using Gaussianized convolution fields}\label{SS:testing}
In order to perform one-sample inference we define $ T^G:S \rightarrow \mathbb{R} $ to be the \textbf{Gaussianized convolution} $ \mathbf{t} $\textbf{-field} sending $ s \in S $ to 
\begin{equation}\label{eq:tstat}
T^G(s) = \frac{1}{\sqrt{N}}\sum_{n = 1}^N Y^G_n(s)\left(\frac{1}{N-1}\sum_{n = 1}^N \left( Y^G_n(s) - \frac{1}{N}\sum_{n = 1}^N Y^G_n(s) \right)^2\right)^{-1/2}.
\end{equation}
Note that $ T^G $ depends on $ N $, but we drop the $ N $ for ease of notation. Then, given a two-sided rejection threshold $ u > 0 $, our inference procedure will conclude that $ \mu_G(s) > 0 $ if $ T^G(s) \geq u $ and that $ \mu_G(s) < 0 $ if $ T^G(s) \leq -u $. Let $ S_0 = \lbrace s \in S: \mu_G(s) = 0 \rbrace$ be the subset of $ S $ on which $ \mu $ is zero. Then we define the \textit{familywise error rate} (FWER) of this rejection procedure to be
\begin{equation*}
	\text{FWER} = \mathbb{P}\left(\sup_{s \in S_0} |T^G(s)| \geq u\right);
\end{equation*}
the probability of a single super-threshold excursion on the set $S_0$.

In Section \ref{S:RFTthresh}, we discuss how to use the GKF to choose a rejection threshold $ u_\alpha $ such that $ \mathbb{P}\big(\sup_{s \in S} |T^G(s)| \geq u_\alpha ~|~ S_0 = S \big) $ is bounded at a level $ \alpha \in (0,1) $, up to the approximation given by the EEC. This choice of threshold provides strong control with respect to the smoothed signal $ \mu $ which holds as for any choice of $ S_0 \subset S $,
\begin{equation*}
	\mathbb{P}\left(\sup_{s \in S_0} |T^G(s)| \geq u\right) \leq \mathbb{P}\left(\sup_{s \in S} |T^G(s)| \geq u \,\middle|\, S_0 = S \right).
\end{equation*}
The right hand side probability can be targetted using RFT to provide a bound on the FWER, see Section \ref{S:RFTthresh} for details\footnote{Note that we write $ \mathbb{P}\left(\sup_{s \in S} |T^G(s)| \geq u \middle| S_0 = S \right) $ to denote the probability given that $ \mu(s) = 0 $  for all $ s \in S$, rather than denoting conditional probability.}.

\begin{remark}
	The Gaussianization transformation does not guarantee that the data is jointly Gaussian and instead targets marginal Gaussianity. However, joint Gaussianity of the test-statistic $ T^G $ is still ensured asymptotically by the CLT. When the data is heavy tailed the test-statistic based on the Gaussianized data converges faster to its limiting distribution (than the test-statistic based on the original data) because the transformed data no longer has heavy tails. This leads to improvements in the accuracy and validity of the RFT inference framework for the small to moderate sample sizes which are typical in fMRI, as we show in Sections \ref{S:ResultsSim} and \ref{SS:RSV}.
\end{remark}



\subsubsection{Implementation of convolution fields}\label{SS:implem}
The difference between the convolution and lattice approaches is clearly shown in Figure \ref{fig:convvslatimage}. Two options are available when calculating convolution fields - either the fields can be evaluated on a higher resolution lattice or alternatively the maximum of the convolution $ t $-field can be found using optimization algorithms. In what follows we perform the later using sequential quadratic programming, \cite{Nocedal2006}. To do so we find the local maxima of the convolution $ t $-field on the $ r = 1 $ lattice $ \mathcal{V}_1 $ and use these locations to initialize the optimization. These approaches are further discussed in Section \ref{AA:convfields}.



\subsection{Controlling the FWER using RFT}\label{SS:EEC}
Computing the EEC is key to controlling the FWER using RFT because for high thresholds $ u $ it closely approximates the expected number of maxima above $ u $ as
\begin{equation}\label{eq:ECapprox}
\mathbb{E}\left[ M_u(T^G) \right] \approx \mathbb{E}\big[ \chi(\mathcal{A}_u(T^G)) \big]
\end{equation}
where $ M_u(T^G) $ is the number of local maxima of $ T^G $ which are greater than or equal to $ u $ and $ \chi(\mathcal{A}_u(T^G)) $ is the Euler characteristic (EC) of the \textbf{excursion set}
\begin{equation*}
\mathcal{A}_u(T^G) = \left\lbrace t \in S: T^G(t) \geq u \right\rbrace.
\end{equation*}
The left hand side of (5) is the \textit{expected Euler characteristic} (EEC).

\subsubsection{The Gaussian Kinematic Formula (GKF)}\label{SS:GKFmaintext}
In order to estimate the EEC we use the GKF \citep{Taylor2006}, the regularity conditions of which are satisfied by convolution fields, see Theorem 1 of Part 1. As such when $ \mu_X = 0, $ the GKF implies that 
\begin{equation}\label{eq:GKF}
\mathbb{E}\big[ \chi(\mathcal{A}_u(T^G)) \big] \approx \sum_{d = 0}^D \mathcal{L}^G_d \rho^{d}_{N-1}(u)\,.
\end{equation}
where the approximation is up to convergence of the test-statistic $ T^G $ to a Gaussian random field. The values $\mathcal{L}^G_0, \dots, \mathcal{L}^G_D  $ are the \textbf{Lipshitz Killing curvatures (LKCs)} of the limiting fields and can be estimated using the component random fields: $ Y^G_1, \dots, Y^G_N $. $ \mathcal{L}^G_0 $ is simply the Euler characteristic of $ S $ \citep{Adler2010}. The $ \rho_{N-1} $ are the \textbf{EC densities} of a $ t $-field with $ N - 1$ degrees of freedom, and can be computed exactly (see \cite{Worsley1992}, Table II for explicit forms).

\subsubsection{Estimating the LKCs}\label{SS:LKCest}
The top two LKCs have well-known closed forms which depend solely on the covariance  $\Lambda^G: S \rightarrow \mathbb{R}^{D \times D}$ of the partial derivatives of the limiting distribution of $ T^G $ (see Section \ref{AA:LKCtheory}). In order to estimate $ \Lambda^G $ we calculate the residual fields
\begin{equation}\label{eq:resid}
R^G_n = \left(Y^G_n - \frac{1}{N}\sum_{n = 1}^N Y^G_n\right)\left(\frac{1}{N-1}\sum_{n = 1}^N \left( Y^G_n - \frac{1}{N}\sum_{n = 1}^N Y^G_n \right)^2\right)^{-1/2},
\end{equation}
$ n = 1, \dots, N $, where operations are performed pointwise.
Then, for each $ s \in S$, let 
\begin{equation}\label{eq:lambdahatconv}
\hat{\Lambda}^G_{ij}(s) = \frac{1}{N-1}\sum_{n = 1}^N \frac{\partial R^G_n(s)}{\partial s_i}\left( \frac{\partial R^G_n(s)}{\partial s_j} - \frac{1}{N}\sum_{m = 1}^N\frac{\partial R^G_m(s)}{\partial s_j} \right)
\end{equation}
for $ i,j = 1, \dots, D, $ and estimate $ \Lambda^G(s) $ using $ \hat{\Lambda}^G(s) \in \mathbb{R}^{D\times D} $: with $ i,j $th entry $ \hat{\Lambda}^G_{ij}(s) $.

We can estimate $ \mathcal{L}^G_D $ and $ \mathcal{L}^G_{D-1} $, at a given added resolution $ r $, by calculating $ \hat{\Lambda}^G $ on the grid $ \mathcal{V}_r $ and using the estimates for the LKCs on a voxel manifold defined in Section \ref{Lambda2LKC}. For $ D \geq 3 $, the remaining LKCs are more difficult to estimate. In this work, for fMRI datasets where $ D = 3 $, we shall use a locally stationary approximation to $ \mathcal{L}^G_1 $ which we introduced in Section 3.5 of Part 1. At higher thresholds (which are the ones that concern us when controlling FWER) $\mathcal{L}^G_{D} $ and $ \mathcal{L}^G_{D-1} $ are more important because of the form of the EC densities. As such, using the locally stationary approximation to $ \mathcal{L}^G_1 $ does not meaningfully affect the ability to estimate the EEC, see Figures \ref{fig:EECplotsFWHM5} and \ref{fig:EECplotsFWHM2}.

\paragraph{Other methods used in fMRI to compute the LKCs}
The implementations of RFT in SPM and FSL use \cite{Kiebel1999}'s lattice based estimate of $ \Lambda = \cov(\nabla Y_1)$ which assumes stationarity. This estimate is also biased at low smoothing bandwidths because of the use of discrete derivatives, see \eqref{eq:DD}. This has long been known, see e.g. \cite{Kiebel1999}'s Figure 3, but is rarely commented on. A second approach discussed in the fMRI literature to estimate $ \Lambda $ \citep{Jenkinson2000} is that of \cite{Forman1995} which has more restrictive assumptions. We provide a discussion of the theory behind the \cite{Kiebel1999} and \cite{Forman1995} approaches in Section \ref{A:latsmo}. We show, in Figures \ref{fig:ECplots} and \ref{fig:EECplotsFWHM2}, that these approaches do not correctly estimate the EEC when applied to fMRI data, regardless of whether the data is transformed. Other LKC estimators, such as the HPE and bHPE approaches of \cite{TelschowHPE}, are compared to the convolution estimates of the LKCs in \cite{Telschow2023FWER}.

\subsubsection{Obtaining the FWER threshold}\label{S:RFTthresh}
Given estimates $ (\hat{\mathcal{L}}^G_d)_{d = 0}^D $ of the LKCs, we can apply the GKF to obtain a FWER threshold. In particular, given an error rate $ \alpha $, we choose the voxelwise threshold $ u_{\alpha} $ to be the highest value $ u \in \mathbb{R} $ such that
\begin{equation}\label{eq:LKCu}
\sum_{d = 0}^D \hat{\mathcal{L}}^G_d \rho^d_T(u) = \frac{\alpha}{2}.
\end{equation}
This ensures that the FWER is approximately controlled to a level $ \alpha $, since
\begin{align*}\label{eq:twotail}
&\mathbb{P}\left( \sup_{s \in S} \left| T^G(s) \right| >  u_{\alpha/2} \,\middle|\, S_0 = S\right) \\
&= 
2\mathbb{P}\left( \sup_{s \in S} T^G(s) \geq u_{\alpha/2}  \,\middle|\, S_0 = S\right) - \mathbb{P}\left(   \sup_{s \in S} T^G(s) \geq u_{\alpha/2}, \min_{s \in S} T^G(s) < -u_{\alpha/2} \,\middle|\, S_0 = S\right)\\
&\leq 2\mathbb{E}\left[ M_{u_{\alpha}}( T^G)\,|\,S_0 = S \right] \approx 2\mathbb{E}\left[ \chi(\mathcal{A}_{u_{\alpha}}( T^G)) \,|\, S_0 = S\right] \approx 2\sum_{d = 0}^D \hat{\mathcal{L}}^G_d \rho^d_T(u_{\alpha}) = \alpha.
\end{align*}
the thresholds used for FWER inference we expect to have 0 or 1 maxima above the threshold with high probability, since we are controlling the expected number of maxima to be $ \alpha $, as such we expect these approximations to be accurate. See Appendix \ref{AA:accuracy} for a full justification for the approximation.

\begin{remark}
	Taking $ u = u_{\alpha} $ in the test defined in Section \ref{SS:testing} provides sign control with respect to $ \mu_G $ because it consists of two one-sided tests at the $ \alpha/2$ level. Thresholds to adjust for one-sided inference can also be obtained using the RFT framework, see Appendix \ref{A:oneside}. 
\end{remark}

\begin{remark}
	This approach provides strong control of the FWER with respect to the smoothed signal $ \mu_G $. However, it also provides information about the original mean
	$ \mu_X = \mathbb{E}(X_1): \mathcal{V} \rightarrow \mathbb{R} $. In particular, under the global
	$ \mu(s) = 0 $ for all $ s \in S $. As such testing using $ T^G $ provides weak control of the FWER with respect to $ \mu_X $. Stronger statements are possible and depend on the support of the kernel
	being applied. To be more precise, given a FWER threshold $ u $, and a point $ s $ such that $ T^G(s) > u_\alpha $, we can conclude with probability at least $ 1-\alpha $ that there must be some $ v \in \mathcal{V} \cap \lbrace t \in \mathbb{R}^D: K(s-t) > 0 \rbrace $ such that $ \mu_X(v) > 0 $. The statements that can be made about the original signal $ \mu_X $ are thus the same as for the original convolution field framework, see the discussion of Part 1 for further details. An uncertainty principle applies in the sense that there is a trade off between the level of applied smoothing and the precision of inference.
\end{remark}

\section{Methods}
In this section we describe the simulations and resting-state validation using an Eklund test that we ran in order to validate our RFT inference framework.

\subsection{Simulations}\label{SS:sims}

%
In our simulations we use heavy tailed marginal distributions in order to simulate the type of non-Gaussianity that can be present in fMRI datasets, see Sections \ref{SS:Gauss} and \ref{SS:nulldist}. In each simulation setting considered, we run $ J \in \mathbb{N} $ simulations and for each $j\in\{1,\ldots, J\}$, $J=5000$, we generate $N\in\mathbb{N}$
i.i.d. lattice random fields $ (X_{nj})_{1\leq n \leq N}: \mathcal{V} \rightarrow \mathbb{R} $.
Here $ (X_{nj}(v))_{v \in \mathcal{V}} $ consists of i.i.d random variables which are $ t_3 $ distributed, and $ \mathcal{V} $ is a central 2D coronal slice of the standard MNI mask. From these we obtain convolution fields with and without Gaussianizing by smoothing with a 3D isotropic Gaussian kernel with given FWHM and calculate corresponding test-statistic fields in each case. This yields test-statistic fields $ T_1, \ldots, T_J $ and Gaussianized test-statistic fields $ T_1^G, \ldots, T_J^G $. 


\subsubsection{Calculating the LKCs on the Gaussianized data}\label{LKCsims}
For each $ N \in \lbrace 20,50,100\rbrace $,
each FWHM $ \in \{2, 3, \ldots, 6\}$ and $ 1 \leq j \leq J $ we compute the estimators $ \hat{\mathcal{L}}^G_{1j} $
and $ \hat{\mathcal{L}}^G_{2j} $ of $ \mathcal{L}^G_1 $ and $ \mathcal{L}^G_2 $ based on the $ N $ Gaussianized fields, with an added resolution of 1, using the LKC estimators discussed in Section \ref{SS:LKCest}. We also calculate the estimates $ \hat{\mathcal{L}}_{1j} $ and $ \hat{\mathcal{L}}_{2j} $ based on the $ N $ original convolution fields. Box plots of the relative bias of these estimates are shown in Figure \ref{fig:LKCplots}. 
 
The ground truth for the LKCs $\mathcal{L}_d$, $d\in\{1,2\}$, are obtained by exploiting the fact
that the data is spatially independent before smoothing. Thus, the $(i,j)$-th entry of the
true $\Lambda^G$ is given by
\begin{equation}\label{eq:trueLKC}
	\Lambda^G_{ij}(s) = \sum_{v\in\mathcal{V}} \left(\tfrac{4\log(2)}{\text{FWHM}^2}\right)^2(s_i-v_i)(s_j-v_j)e^{-\frac{4\log(2)\Vert s-v\Vert^2}{\text{FWHM}^2}}\,.
\end{equation}
Using \ref{eq:trueLKC} the true LKCs over the voxel manifold are calculated using an added resolution of 21 by using \eqref{eq:trueLKC} in the formulas for $ \mathcal{L}_{D-1} $ and $ \mathcal{L}_D $ in Section \ref{Lambda2LKC}. Note that in this example $ X_{nj}(v) $ is i.i.d. across voxels and so $ \Lambda^G = \Lambda $ and so $ \mathcal{L}^G_d = \mathcal{L}_d $ for $ d \in \lbrace 0, 1 \rbrace. $ This allows us to compare the LKC estimates on the same scale in Figure \ref{fig:LKCplots}. 


\subsubsection{Validating Gaussianized RFT based FWER inference}\label{sim:GaussFWER}
Taking $ \alpha = 0.05, $ for each $ 1 \leq j \leq J $ we apply our RFT inference procedure to obtain $ \alpha$-level FWER thresholds $ \hat{u}_{\alpha, j}$ based on the convolution fields and $ \hat{u}_{\alpha, j}^G $ based on the Gaussianized convolution fields, as discussed in Section \ref{S:RFTthresh}. We then estimate the FWER provided by RFT on the untransformed test-statistics by
\begin{equation}\label{eq:FWERconv}
\frac{1}{J} \sum_{j = 1}^J 1[\sup_{s \in \mathcal{V}_r} |T_{j}(s)| > \hat{u}_{\alpha, j}]
\end{equation}
for added resolution $ r \in \lbrace 0, \infty \rbrace$, where $  \mathcal{V}_\infty = S. $ Since $ \mathcal{V}_0 = \mathcal{V} $, taking $ r = 0 $ in \eqref{eq:FWERconv} allows us to estimate the FWER that results from evaluating the test-statistics on the original lattice while $ r = \infty $ provides the coverage obtained when using the convolution test-statistic. Likewise for the Gaussianized test-statistics we estimate FWER provided by RFT in the different settings (i.e.  $ r \in \lbrace 0, \infty \rbrace$) by calculating
\begin{equation*}
\frac{1}{J} \sum_{j = 1}^J 1[\sup_{s \in \mathcal{V}_r} |T^G_{j}(s)| >  \hat{u}_{\alpha, j}^G]
\end{equation*}
We repeat this procedure for $ N \in \lbrace 20,50,100\rbrace $ and an applied FWHM of $ 2, 3, 4, 5, 6$ and display the results in Figure \ref{fig:FWHMgauss}.





\subsection{Resting State Validation Approach}\label{SS:UKB}
In order to validate whether our method correctly controls the FWER in practice we apply the Eklund test introduced in \cite{Eklund2016} and \cite{Eklund2018} to resting state data from the UK biobank. This procedure regresses fake task designs on resting state data and uses the resulting contrast maps to provide highly realistic noise on which we can validate the false positive rate of our methods.

\subsubsection{Pre-processing the resting state data}\label{SS:rsprocess}
In order to perform our validations we take the approach of \cite{Eklund2016} but instead use resting state data from 7000 subjects from the UK BioBank. For each subject we have a total of 490 T2*-weighted blood-oxygen level-dependent (BOLD) echo planar resting state images [TR = $ 0.735 $s, TE = $ 39 $ ms, FA = $ 52^{\circ} $, $ 2.4 $mm$ ^3 $ isotropic voxels in $ 88 \times88\times64 $ matrix, $ \times 8 $ multislice acceleration]. We pre-processed the data for each subject using FSL \citep{Jenkinson2012} and used a block design at the first level consisting of alternating blocks of 20 time points to obtain first level contrast images. This pre-processing included convolution of the design with a haemodynamic response function and correction for motion correction.

\begin{figure}[h]
	\begin{center}
		\includegraphics[width=0.37\textwidth]{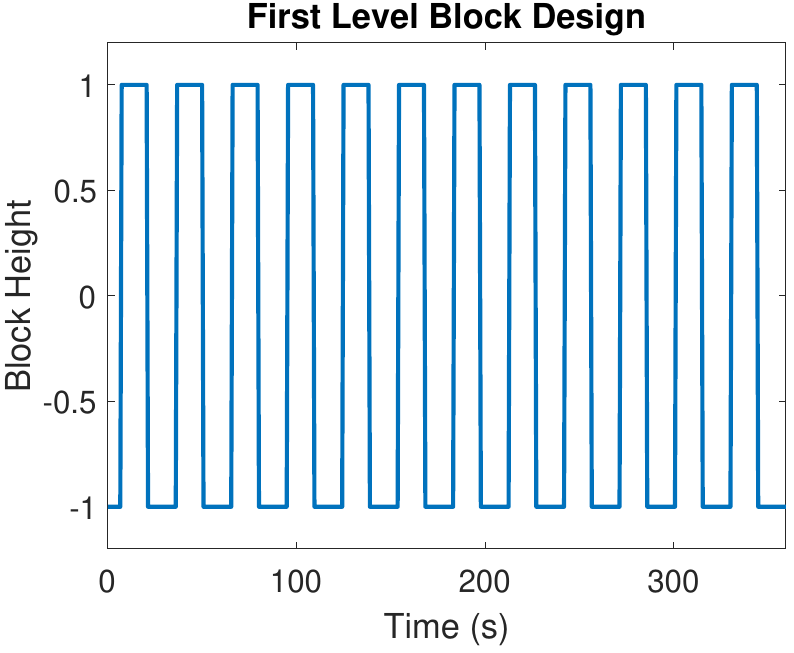}
		\hspace{1cm}
		\includegraphics[width=0.4\textwidth]{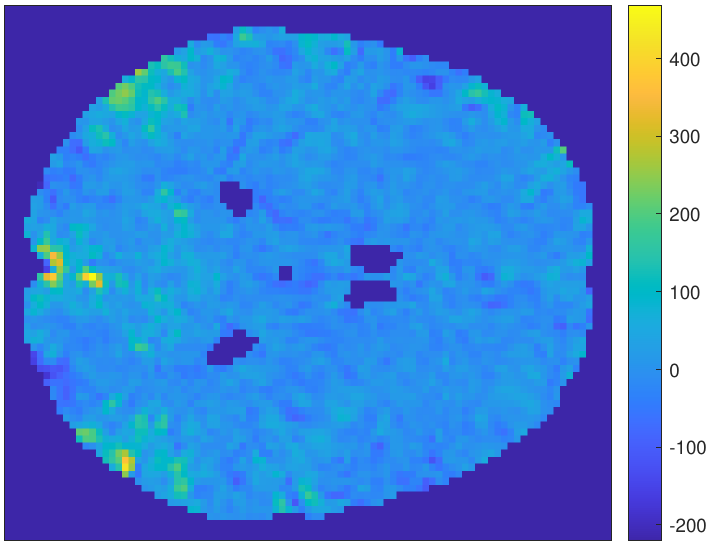}
	\end{center}
	\caption{The first level block design (before convolution with the haemodynamic response function). The left plot shows one of the random block designs that we used at the first level. The other random blocks used are randomly shifted versions of this design. The right image displays a coronal slice through the brain of the COPE image of one of the subjects that has been processed through FSL using this block design before the data has been smoothed.}\label{fig:blockdesign}
\end{figure}

Specifically, for each subject $ 1 \leq l \leq 7000 $, let $ Z_l$ be the time series of 490 images in the subject's native space after first level pre-processing. Moreover, define $\mathcal{B}_i \subset [0,489] $ to be a regular block design with a block length of $ 20 $ and a uniform phase shift, drawn independently over subjects. An example such block design is shown in Figure \ref{fig:blockdesign}. The first level contrast maps that result from using these designs are,
\begin{equation*}
	\sum_{i \in\mathcal{B}_l} Z_l(i) - \sum_{k \in [0,489]\setminus\mathcal{B}_l } Z_l(k),
\end{equation*} 
which are defined within the native space for each subject. We transform these contrast maps to 2 mm MNI space using nonlinear warping determined by the T1 image and an affine registration of the T2* to the T1 image.

Masks were obtained for each subject using the standard FSL pipeline, which identifies brain voxels based on mean brain functional intensities and morphological properties. The mask for each subject will thus be different. In order to compare the Euler characteristic curves across samples, for the purpose of the validations that we describe in the following sections, we use intersection mask taken across all 7000 subjects as is standard in neuroimaging. As such, after pre-processing we obtain 3D contrast maps $ A_1, \dots, A_{7000} $ defined on the intersection mask. 

\cite{Eklund2016} used a fixed block design in their resting state data validations. They assumed that the resulting contrast-maps were mean zero, as there is no consistent localized activation across subjects. However this assumption may not hold in practice \citep{Slotnick2017a}. In particular there is no guarantee that the resulting contrast maps are mean-zero because there could be systematic effects at beginning or end of the scan which could correlate with a fixed block. In order to ensure that our contrast maps are mean zero and thus suitable for performing validations, as described above, we instead use a block design for each subject with a random phase shift. Since there is no reason that the design should correlate with the resting state time series, the resulting contrast maps are mean-zero. 

\subsubsection{Subsampling the resting-state contrast maps}
Given a number of subjects $ N $, following \cite{Eklund2016}, we draw samples of size $ N $ from $ A_1, \dots, A_{7000} $ with replacement. We do this $ J = 5000 $ times and thus for $ 1 \leq j \leq J $ obtain a sample $ (X_{nj})_{1\leq n \leq N} $ where $ X_{nj} = A_{d_n(j)}$ and $(d_n(j))_{1\leq n \leq N} \subset \lbrace 1 \dots, 7000 \rbrace $ indexes the $ j $th draw. Within the $ j $th sample we obtain convolution fields and Gaussianized convolution fields, as in Section \ref{SS:sims}, by smoothing the contrast maps within the sample with a 3D isotropic Gaussian kernel with given FWHM. We also obtain corresponding test-statistics $ T_j $ and $ T_j^G $. Here the Gaussianization is separately performed within each subsample. 

In these validations we apply a smoothing of $ 1,2, \dots, 6 $ FWHM (in voxels) and take samples sizes $ N \in \lbrace 10,20,50\rbrace$. In each setting we perform the above resampling procedure, taking the draws to be independent across all settings considered.

%

\subsubsection{Establishing the validity of the GKF using the empirical Euler characteristic curves}
%

In this section we discuss how we perform an Eklund test to determine the validity of the GKF, in terms of its ability to estimate the EEC. 

For each setting considered (i.e. each given $ N $ and choice of applied smoothing), for $1 \leq j \leq J$, we define the Euler characteristic (EC) curve $ \chi_j: \mathbb{R} \rightarrow \mathbb{R} $ taking $ u \in \mathbb{R} $ to
\begin{equation*}
\chi_j(u)  = \chi(\mathcal{A}_u(T^G_j)),
\end{equation*}
where $ \chi $ and $ \mathcal{A}_u $ are defined as in Section \ref{SS:EEC}.
where for our domain we use the voxel manifold corresponding to the mask obtained from the intersection of the masks across all 7000 subjects. In order to validate the theory, we compare to the \textbf{empirical EEC curve},
\begin{equation}\label{eq:ECaverage}
u \mapsto \frac{1}{J} \sum_{j = 1}^J \chi_j(u).
\end{equation}
obtained from all the $ J $ samples. 

We use \eqref{eq:ECaverage} as a ground truth against which to test the ability of the GKF to estimate the EEC. To do so we can compare the EEC curves obtained from the drawn samples of the data. In each setting, for $ 1 \leq j \leq J $, we compute estimates $(\hat{\mathcal{L}}^G_{d,j})_{0\leq d \leq D}$ based on the $ N $ images in the $ j $th sample as described in Section \ref{SS:LKCest}. These LKC estimates are different for each sub-sample. As such in order to compare the empirical EEC curve \eqref{eq:ECaverage} to the theoretical prediction from the GKF we average the LKC estimates across all 5000 samples and then use the resulting LKCs to calculate the EEC using \eqref{eq:GKF}. We then compute the curves
\begin{equation}\label{eq:EEEC}
u \mapsto \frac{1}{J}\sum_{j = 1}^J\sum_{d = 0}^D \hat{\mathcal{L}}^G_{d,j} \rho^{d}_{N-1}(u)
\end{equation}
and compare them to the empirical expected EC curve \eqref{eq:ECaverage}. Here $\rho^d_{\nu}$ is the EC density of the $ t $-statistic with $ \nu = N-1 $. By comparing the curves \eqref{eq:ECaverage} and \eqref{eq:EEEC} we are able to compare the theoretical estimates to the actual EEC of the excursion set at each threshold. This comparison allows us to determine the accuracy and bias of our estimates of the EC curves. We repeat this procedure for the test-statistics $ (T_j)_{j = 1}^{J} $, instead using the LKC estimates $ (\hat{\mathcal{L}}_{dj})_{d = 1}^D$, calculated using the original convolution fields.

We do the same but with the Kiebel and Forman estimates of the LKCs \citep{Kiebel1999, Forman1995}, defined in Sections \ref{SS:remmet} and \ref{SS:FormanFWHM}. As such in total we obtain six versions of the curve shown in \eqref{eq:EEEC}, one for each method of computing the LKCs for both the original and transformed data. We repeat this for $ N \in \left\lbrace 10,20,50 \right\rbrace $ and for an applied FWHM of $2$ and $ 5 $ voxels. The results are shown in Section \ref{SS:EEEC}.


\subsubsection{Validating the control of the FWER}\label{rsFWER}
We apply our RFT pipeline to this pre-processed data to evaluate whether we obtain the nominal FWER. For $ 1 \leq j \leq J $, we calculate the LKCs and use these to obtain two-tailed $ \alpha $-thresholds ($ \hat{u}_{\alpha, j} $ and $ \hat{u}^G_{\alpha, j} $) for $ \alpha \in \left\lbrace 0.05, 0.01 \right\rbrace $. We estimate the FWER in each case for added resolution $ r \in \lbrace 0, 1, \infty\rbrace $ by computing 
\begin{equation}\label{eq:rsfwer}
\frac{1}{J} \sum_{j = 1}^J \mathds{1}\Big[\sup_{s \in \mathcal{V}_r} |T_{j}(s)| > \hat{u}_{\alpha, j}\Big] \quad \text{ and } \quad \frac{1}{J} \sum_{j = 1}^J \mathds{1}\Big[\sup_{s \in \mathcal{V}_r} |T^G_{j}(s)| > \hat{u}^G_{\alpha, j}\Big]
\end{equation}
for the original fields and the Gaussianized fields respectively. Here the thresholds are obtained using RFT inference on the corresponding fields as described in Section \ref{sim:GaussFWER}. We perform this validation for $ N \in \left\lbrace 10, 20, 50 \right\rbrace $, and applied smoothing levels of $ \left\lbrace 2,3,4,5,6 \right\rbrace $ FWHM per voxel. The results are shown in Section \ref{SS:fwer}.

\begin{remark}\label{rmk:dependence}
	Our dataset consists of 7000 subjects. This is substantially larger than the datasets used in the validations in \cite{Eklund2016,Eklund2018} which consisted of at most 198 subjects. Using a larger number of subjects is advantageous because it means that the resamples of $ N $ subjects from the dataset are more independent (because they are much less likely to overlap) and thus better model different fMRI studies. As a result our resamples are able to provide a more precise sample of the null distribution.
\end{remark}

In order to account for the dependence in their draws \cite{Eklund2016,Eklund2018} generated the confidence bands for their FWER estimates using stationary Gaussian simulations (for further details see the supplementary material of \cite{Eklund2016}). However the lack of stationarity \cite{Eklund2016} and Gaussianity (see Section \ref{SS:Gauss}) in fMRI data means that it is difficult to ensure the accuracy of these confidence bands. In our figures we instead treat the draws as independent in order to calculate confidence bands for the coverage rates using the CLT approximation to the binomial distribution. This approach was not feasible for \cite{Eklund2016} since they had to account for dependence between their resamples of size $N$ as their datasets consisted of at most 198 subjects. However, it is reasonable in our setting where we are drawing samples of size at most 50 from a pool of 7000 subjects, see Remark \ref{rmk:dependence}.

\section{Results - Simulated Data}\label{S:ResultsSim}

\subsection{Simulations validating the RFT inference framework}

\subsubsection{LKC estimation}\label{SS:simlkcresults}
\begin{figure}[h!]
	\begin{center}
	\begin{subfigure}[b]{0.48\textwidth}
		\centering
		\includegraphics[width=\textwidth]{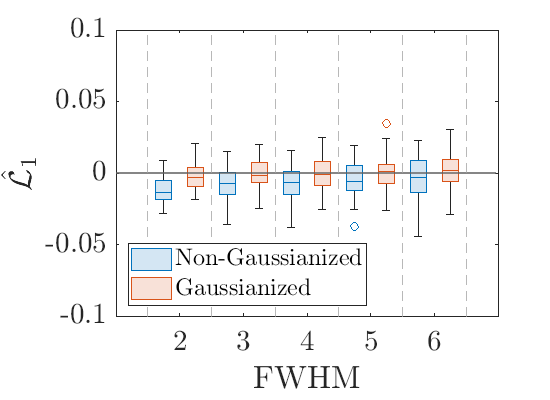}
	\end{subfigure}
	\begin{subfigure}[b]{0.48\textwidth}
		\centering
		\includegraphics[width=\textwidth]{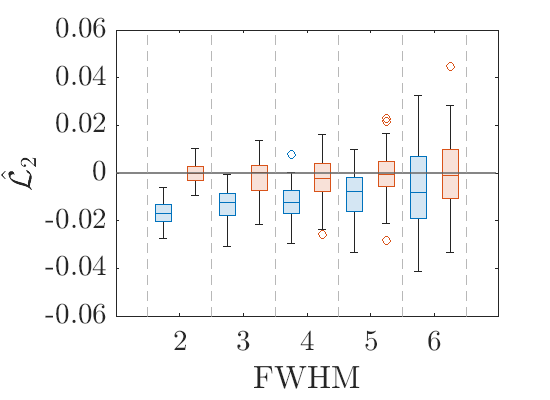}
	\end{subfigure}
\end{center}
	\caption{The relative bias
			 of the LKC estimates for different levels of applied smoothing, when using a sample of 50 convolution fields obtained by smoothing i.i.d. $ t^3  $ noise. Results for the original data are shown in blue while the results for the Gaussianized data shown in red.}\label{fig:LKCplots}
\end{figure}

In Figure \ref{fig:LKCplots} the relative bias of the LKC estimates, obtained as described in Section \ref{LKCsims} is shown. There are two main conclusions from these simulations.
Firstly, the LKC estimators based on the original untransformed data have not yet converged to the LKCs of the limiting field due to heavy tails and are thus biased (blue box plots). The validity of the estimators is significantly improved after applying our Gaussianization procedure to the data before calculation of the LKCs: the resulting estimator (shown in the red box plots) seems to be almost unbiased. Secondly, this was achieved with, consistent with our Gaussian simulations from Part 1, with added resolution 1 for FWHM $ \geq 2 $. For lower levels of applied smoothing it would probably be necessary to add resolution in order to obtain better estimates (see, e.g., \cite{Telschow2023FWER}'s Figures 6 and 7).

Notably the relative bias for $\mathcal{L}_2$, which is more important for voxelwise inference than $\mathcal{L}_1$, is lower than the relative bias for $\mathcal{L}_1$. This is evidence that the CLT for the Gaussianized data is in effect at these sample sizes. Appendix \ref{app:Figures} contains additional simulations illustrating the robustness of our LKC estimation to other non-Gaussian random fields.

\subsubsection{FWER control}
In this section we demonstrate that our Gaussianization procedure removes the conservativeness of voxelwise RFT when
applied to heavy tailed data.

\begin{figure}[h]
	\begin{center}
		\includegraphics[width=0.32\textwidth]{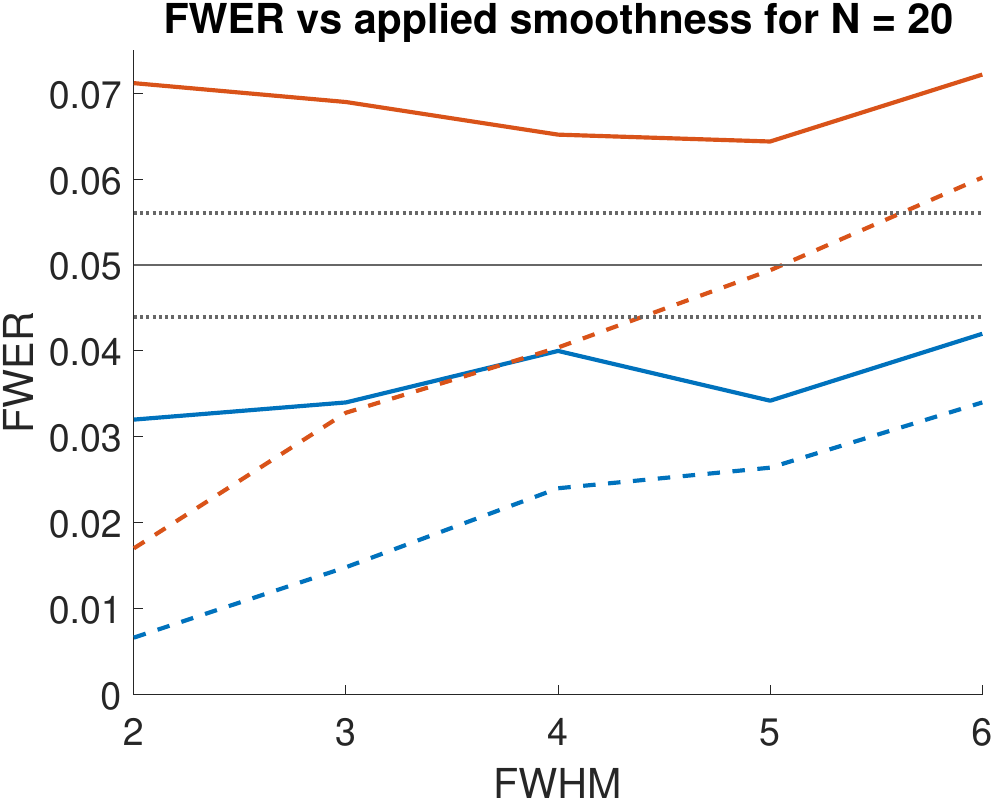}
		\includegraphics[width=0.32\textwidth]{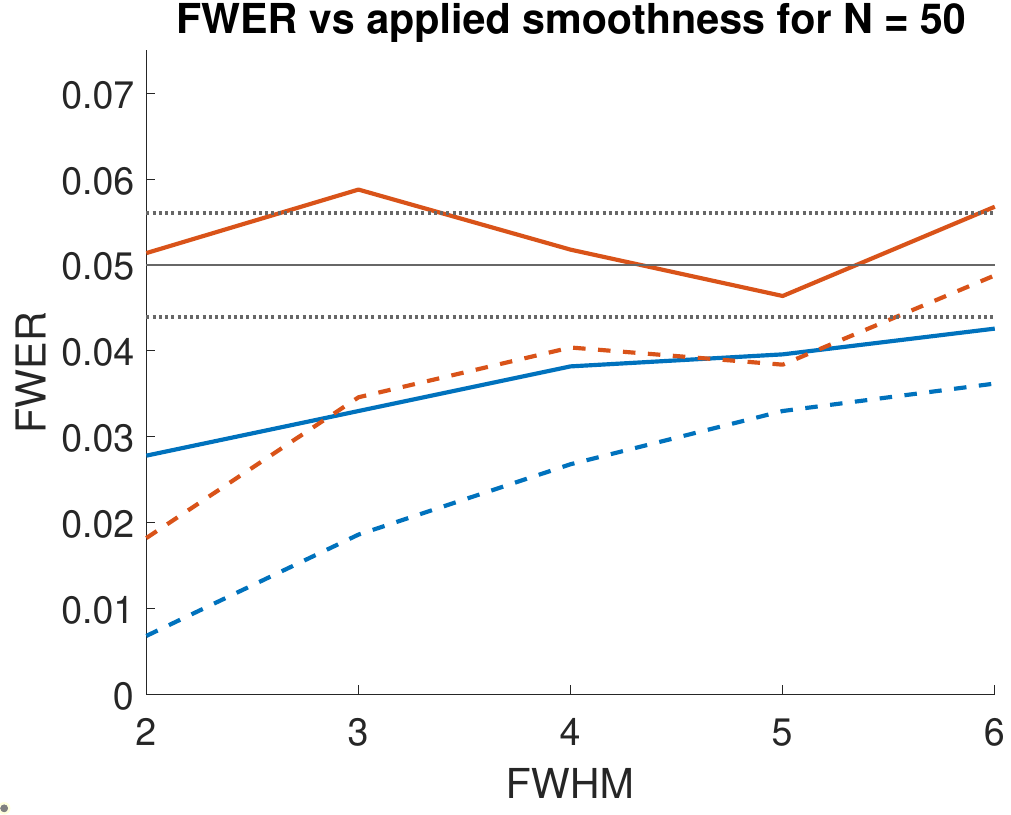}	\includegraphics[width=0.32\textwidth]{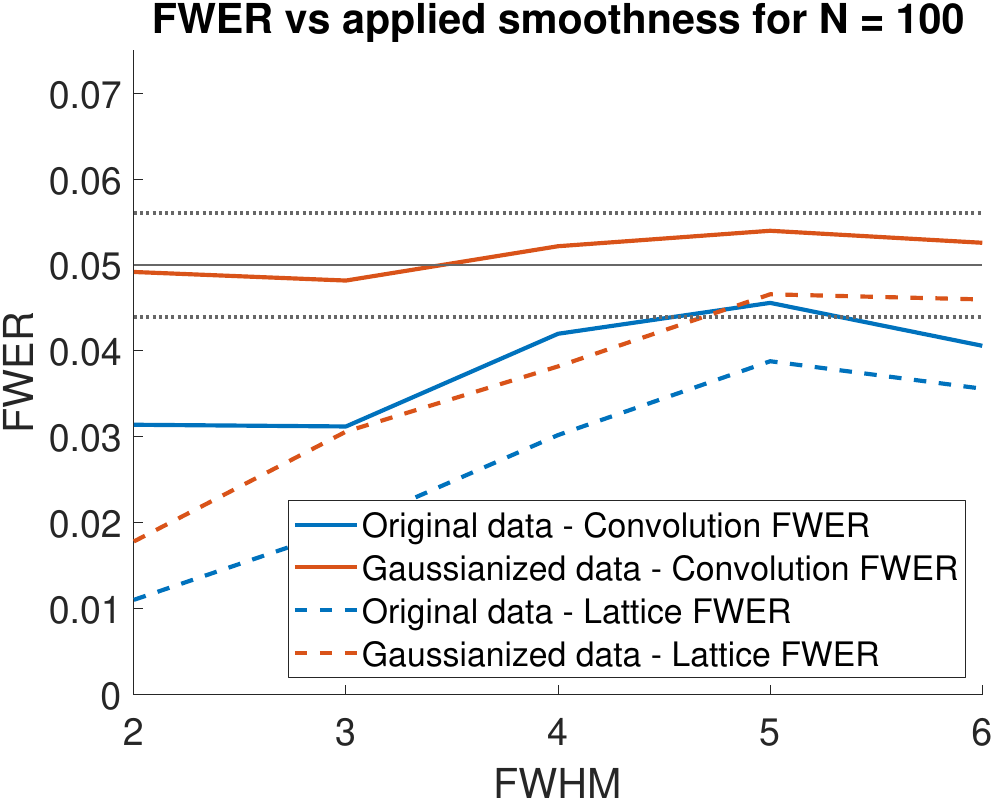}
	\end{center}
	\caption{Plotting the FWER against the applied smoothness for $ N \in \left\lbrace 20,50,100 \right\rbrace $ when applying the RFT methods to the original (marginal $ t_3 $) data and to the Gaussianized data. The methods are conservative when applied to the original data even when the number of subjects is quite large. When the data is Gaussianized and convolution fields are used the FWER (solid red) is accurately controlled to the nominal level once the number of subjects is sufficiently high. If the traditional lattice maximum (dashed red) is used rather than the convolution maximum then a high level of smoothness is required before the methods attain the nominal FWER. For the original data (shown in blue), as the smoothness increases the FWER becomes closer to the nominal rate. This is because more averaging is involved and so the CLT comes into affect more quickly.}\label{fig:FWHMgauss}
\end{figure}

\begin{figure}[h]
	\begin{center}
		\includegraphics[width=0.46\textwidth]{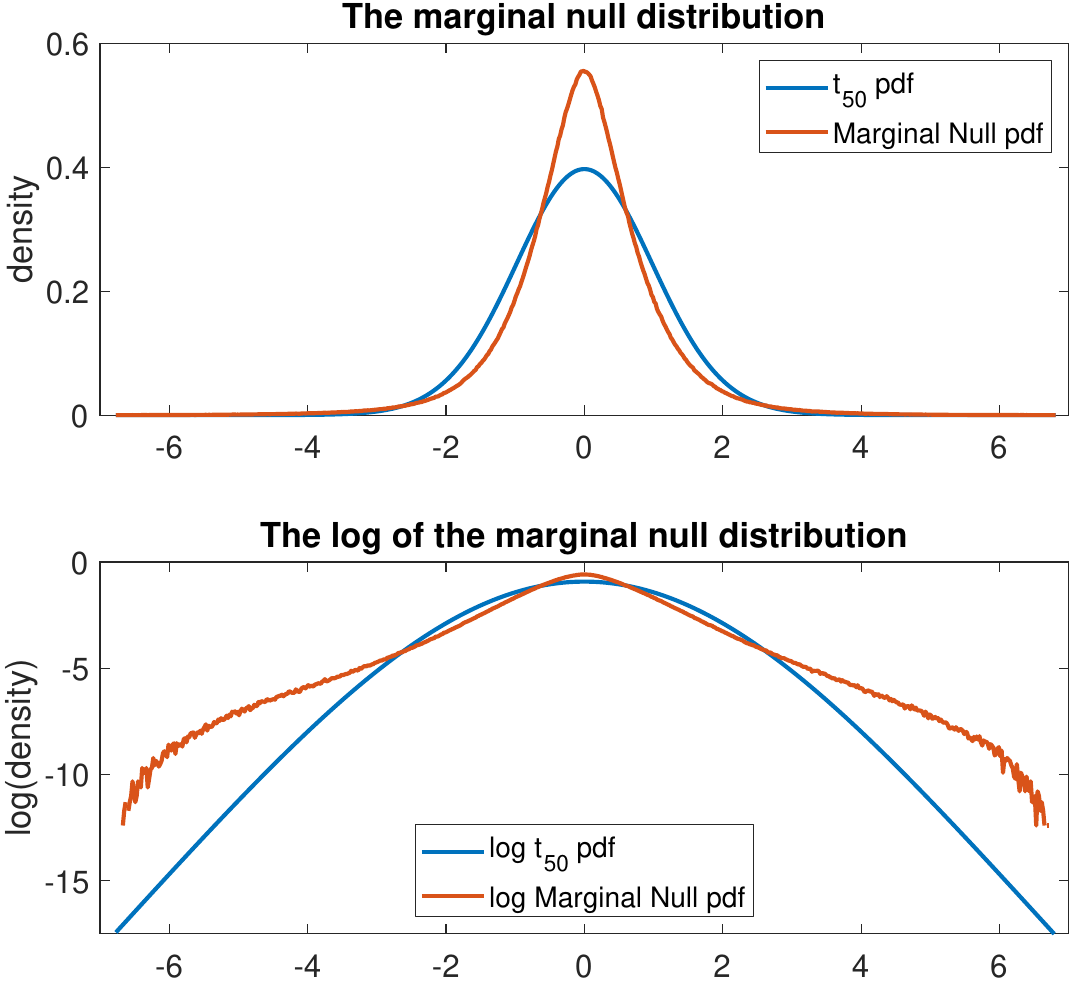}
		\hspace{1cm}\includegraphics[width=0.46\textwidth]{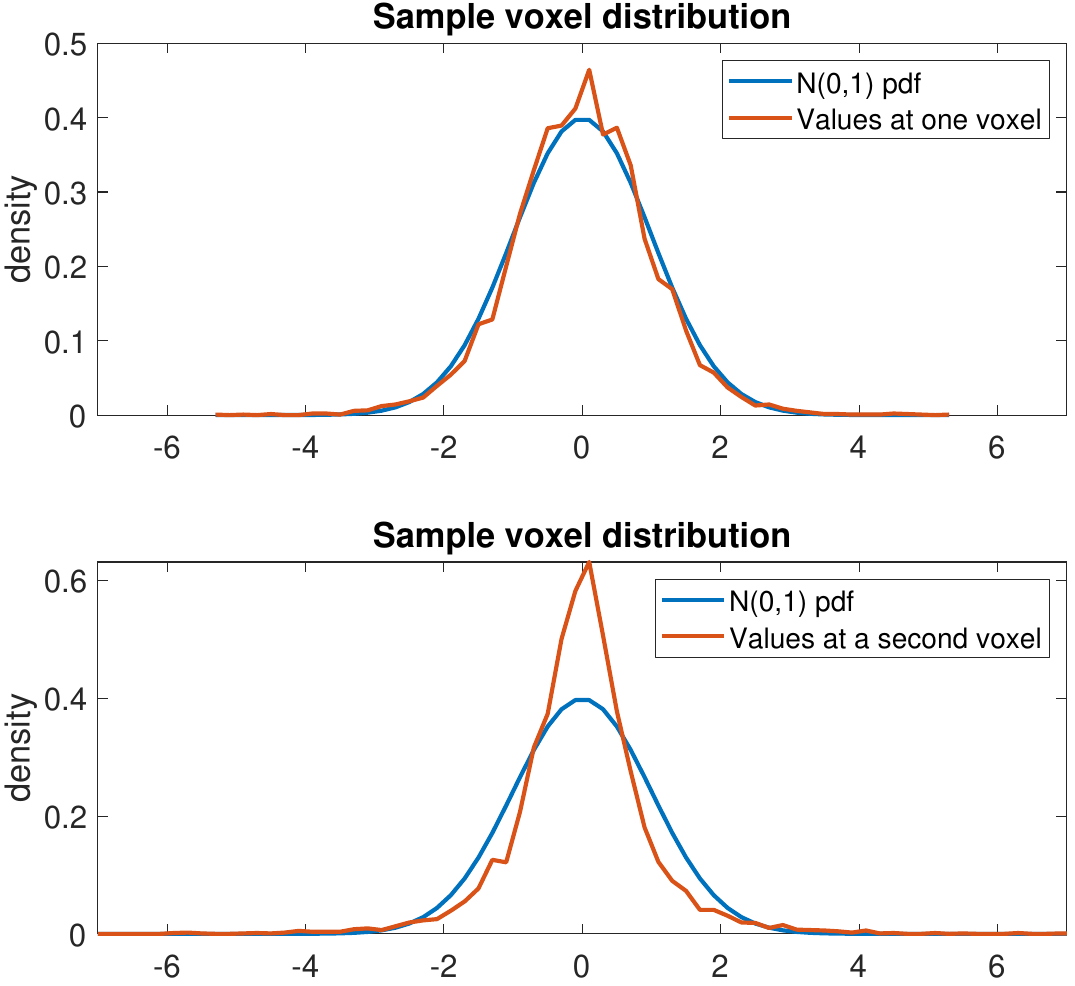}
	\end{center}
	\caption{The marginal distributions of the resting state data processed with the randomized block design. In the plots on the left we plot the histogram of the marginal null distribution (the density and its log) for a sample of 50 randomly selected subjects (over the values for all subjects and voxels in the 50 subject intersection mask) calculated using (\ref{eq:marginalnull}) against the $ t_{50} $ pdf. On the right we plot two histograms of the observed values at two different voxels, calculated over all 7000 subjects (and scaled to have variance 1) against the pdf of the normal distribution. As can be seen from these plots the degree of non-Gaussianity can vary across voxels.}\label{fig:marginalnull}
\end{figure}

The results for the setting described in Section \ref{sim:GaussFWER} are shown in Figure \ref{fig:FWHMgauss} for different sample sizes $N$. We compare the FWER that results from applying RFT on the original lattice and on the convolution random field. This is done for the original data and the Gaussianized data. From these plots we see that without Gaussianization the FWER control is conservative for heavy-tailed data even if convolution fields are used. Once we Gaussianize the FWER is slightly anti-conservative for $ N = 20 $, close to the nominal level for $ N = 50, 100$ and controlled to the nominal level at $ N = 100 $ for all smoothness levels.

For the original data the EEC is incorrectly estimated (see Figure \ref{fig:ECplots}) and the FWER control is conservative as a result. This occurs because the data is not Gaussian and so the GKF does not hold. However, applying the RFT framework to the Gaussianized data and using convolution fields, we can accurately estimate the EEC and obtain valid and accurate FWER control given sufficient numbers of subjects. For lower numbers of subjects the error rate, after Gaussianization, is slightly inflated as the test-statistics have not yet converged. 

This occurs for several reasons. Firstly, under the transformation the heavy tailed nature of the data is decreased. However, after Gaussianization the data is not perfectly marginally Gaussian distributed for low numbers of subjects because subtracting the empirical mean is not equivalent to assuming that the data is mean zero; this problem decreases as the number of subjects increases. Secondly while the data (under the null) is transformed to be marginally close to Gaussian it is not jointly Gaussian. Thirdly for low number of subjects the Gaussianization induces some dependence between subjects meaning that the resulting test-statistics are not $ t $-fields. These issues are negligible asymptotically due to consistent estimation of the mean and the CLT.


%
%
%

\section{Results - Resting State Validation}\label{SS:RSV}
In this section we validate our voxelwise RFT framework using UK BioBank resting state fMRI data, analysed as if it were task data. We show that the GKF fails when applied to the original (untransformed) data, leading to conservative error rates. Instead when the data is Gaussianized we demonstrate that the GKF can be used to correctly calculate the EEC and provide improved inference.

\subsection{The null distribution of the pre-processed resting state data}\label{SS:nulldist}
We first study the distribution of the null distribution of the resting data pre-processed as described in Section \ref{SS:rsprocess}. To do so we have plotted the marginal null distribution of the resting state data (demeaned and standardized) across all voxels and subjects as well as the distribution at some sample voxels. The resulting plots demonstrate that the resulting noise can be highly non-Gaussian, in particular having a distribution that has a lot of weight at the centre and very heavy tails. The histograms of the data at each voxel illustrate that the level of non-Gaussianity varies throughout the brain. 

%
%

\subsection{Empirical vs Expected Euler characteristic}\label{SS:EEEC}

\begin{figure}[h!]
	\begin{center}
		\begin{subfigure}[b]{0.3\textwidth}
			\centering
			\includegraphics[width=\textwidth]{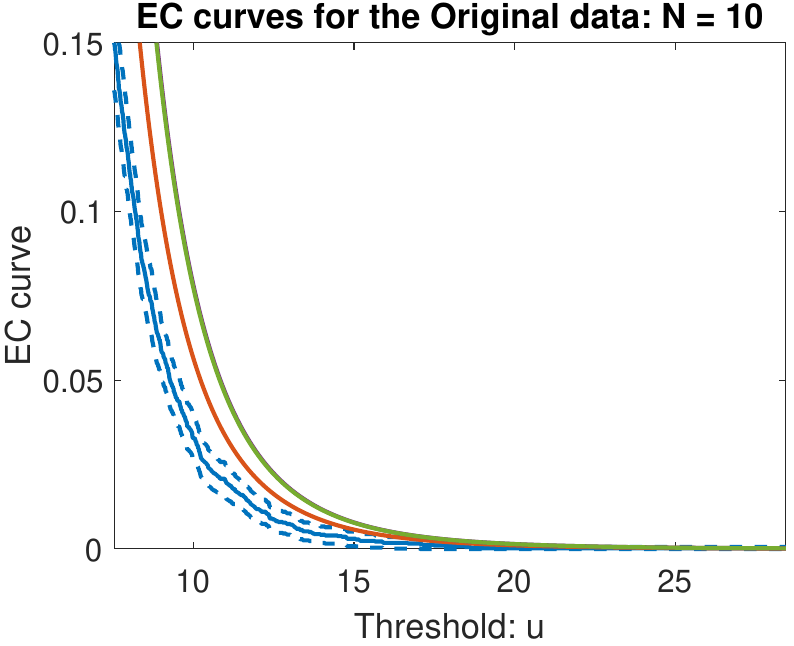}
		\end{subfigure}
		\begin{subfigure}[b]{0.3\textwidth}
			\centering
			\includegraphics[width=\textwidth]{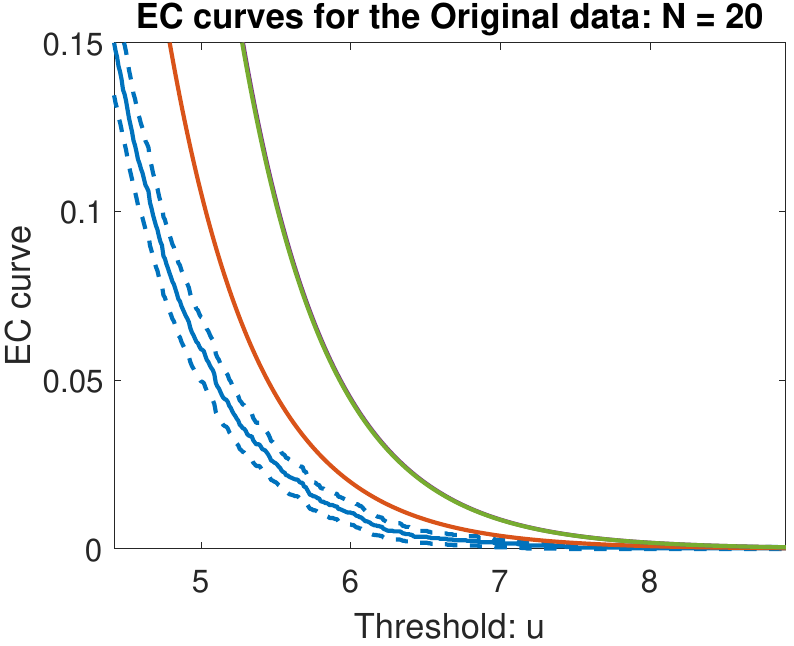}
		\end{subfigure}
		\begin{subfigure}[b]{0.3\textwidth}
			\centering
			\includegraphics[width=\textwidth]{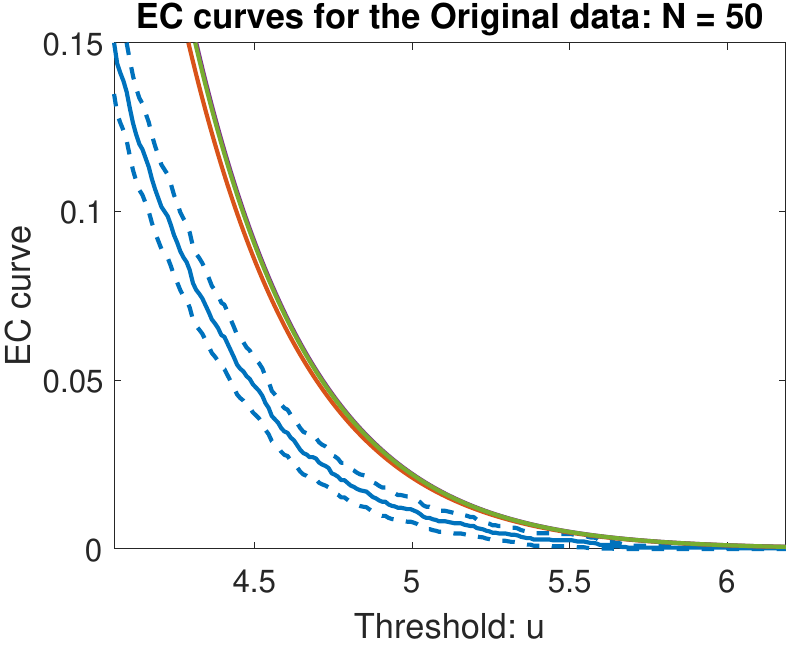}
		\end{subfigure}
	\end{center}
	\vskip\baselineskip
	\begin{center}
		\begin{subfigure}[b]{0.3\textwidth}
			\centering
			\includegraphics[width=\textwidth]{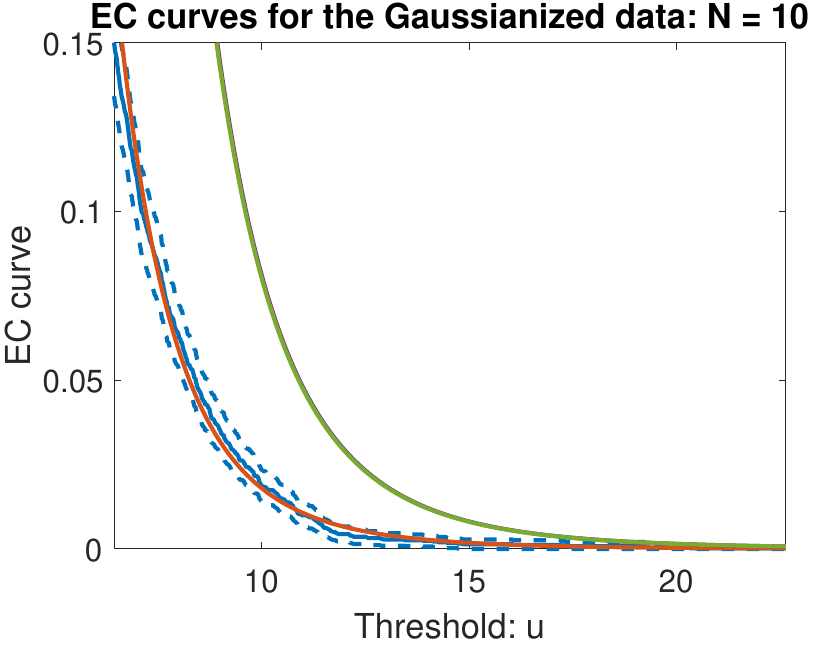}
		\end{subfigure}
		\begin{subfigure}[b]{0.3\textwidth}
			\centering
			\includegraphics[width=\textwidth]{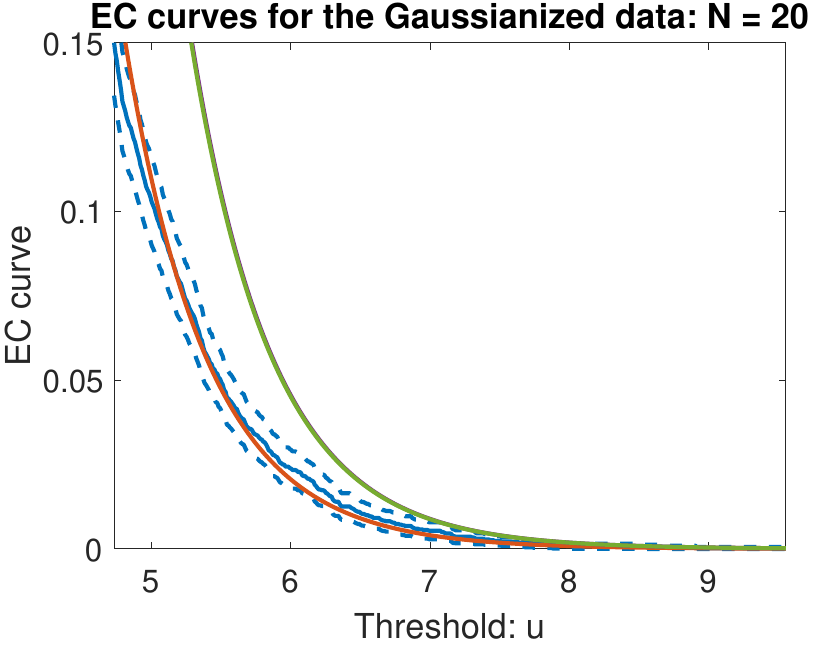}
		\end{subfigure}
		\begin{subfigure}[b]{0.3\textwidth}
			\centering
			\includegraphics[width=\textwidth]{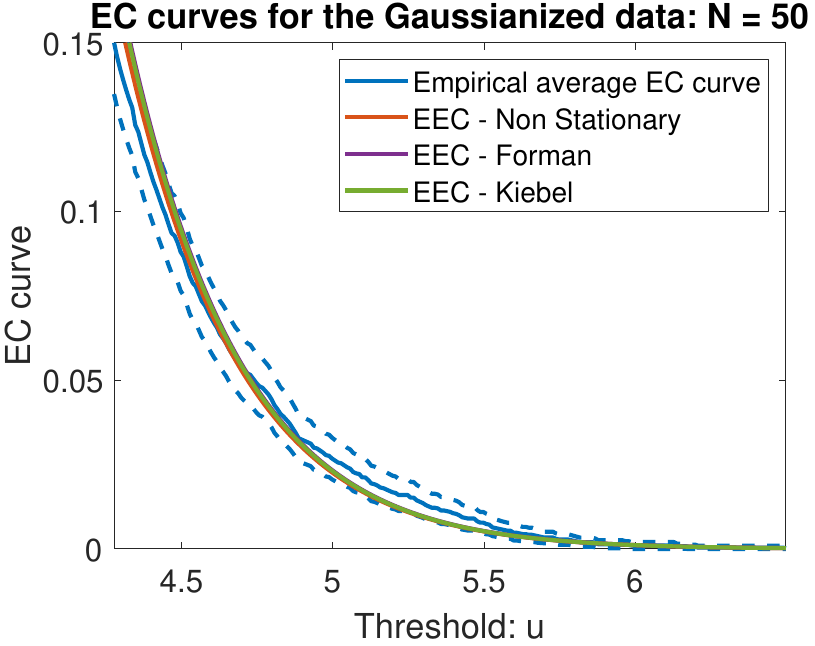}
		\end{subfigure}
	\end{center}	
	\caption{Comparing the expected and empirical tail EC curves in different settings using the resting state data for an applied smoothing of $ 5 $ FWHM per voxel. The results for the Gaussianized data (top row) and for the original data (bottom row) are noticeably different. For the Gaussianized data, the non-stationary expected EC curve is close to the empirical average curve and always lies within the 95\% error bars (calculated using the CLT). For the original data the EEC is overestimated in all cases, which leads to conservative inference. Note that the Kiebel and Forman EC curves are typically so close to each other that they are indistinguishable and provide over estimates other than for $ N = 50 $.}\label{fig:EECplotsFWHM5}
\end{figure}

As described in Section \ref{SS:UKB}, we can compare theoretical (shown in red) and empirical EC curves (shown in blue) to see how well the GKF is estimating the EEC in practice. We have plotted the upper tails of these curves - which are the sections of the curve most relevant for FWER control - for applied smoothnesses of 2 and 5 FWHM per voxel in Figures \ref{fig:EECplotsFWHM5} and \ref{fig:EECplotsFWHM2}. As can be seen from the plots the GKF gives a close estimate of the EEC once the data is Gaussianized. It lies within the confidence bands at all points for FWHM = 5 and at all except the lowest thresholds for FWHM = 2. Without Gaussianization the EEC is instead over-estimated at all thresholds. We perform a similar analysis for the simulated data setting with an applied FWHM of 4 voxels: see Figure \ref{fig:ECplots}. The results of this are similar to the  resting state data.

The stationary methods (shown in green and purple) typically overestimate the EEC curves. However their performance is somewhat erratic and in some settings, once the data is Gaussianized, they do not perform badly. The GKF is a more reliable estimator and works well after the data is transformed. As such, we can conclude that the noise is both non-stationary and non-Gaussian. Over-estimating the EEC leads to conservativeness as it results in a larger than necessary threshold $ u_{\alpha} $. These plots thus indicate that a failure to estimate the EEC is a large cause of the conservativeness of voxelwise RFT.


\subsection{FWER error rate}\label{SS:fwer}
\begin{figure}[h!]
	\begin{center}
		\begin{subfigure}[b]{0.3\textwidth}
			\centering
			\includegraphics[width=\textwidth]{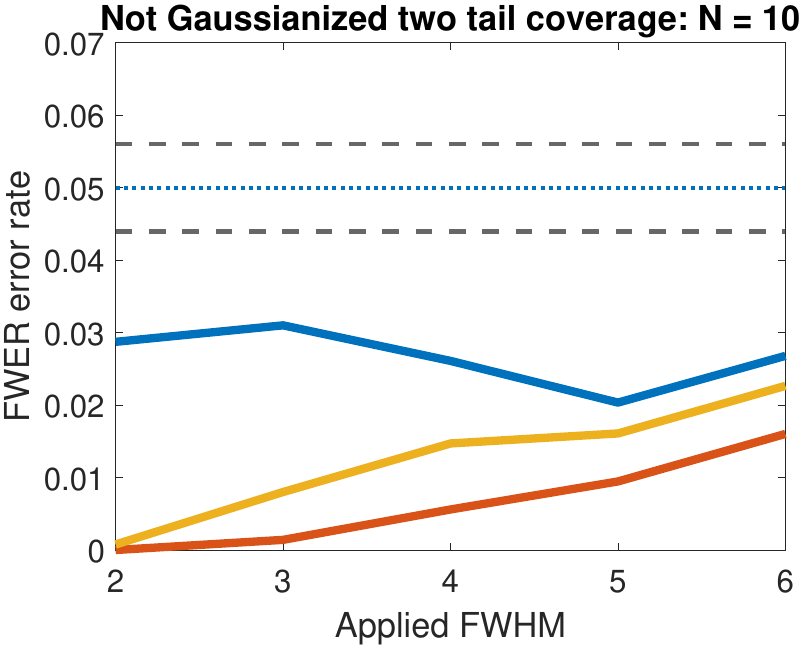}
		\end{subfigure}
		\begin{subfigure}[b]{0.3\textwidth}
			\centering
			\includegraphics[width=\textwidth]{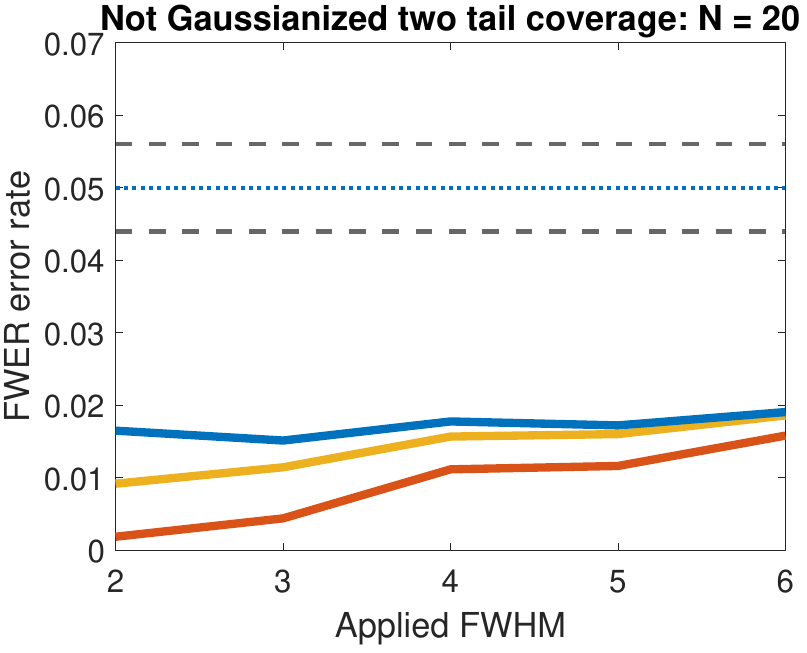}
		\end{subfigure}
		\begin{subfigure}[b]{0.3\textwidth}
			\centering
			\includegraphics[width=\textwidth]{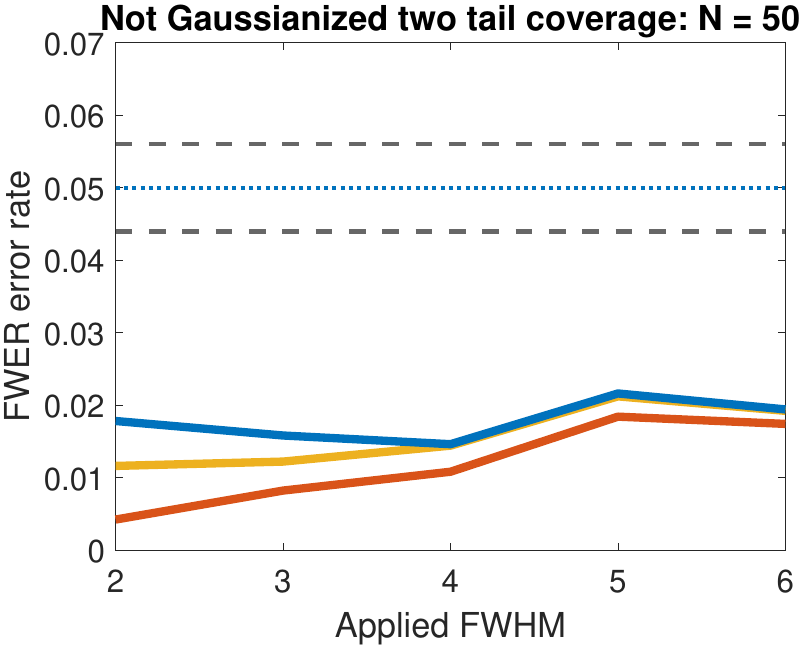}
		\end{subfigure}
	\end{center}
	\vskip\baselineskip
	\begin{center}
		\begin{subfigure}[b]{0.3\textwidth}
			\centering
			\includegraphics[width=\textwidth]{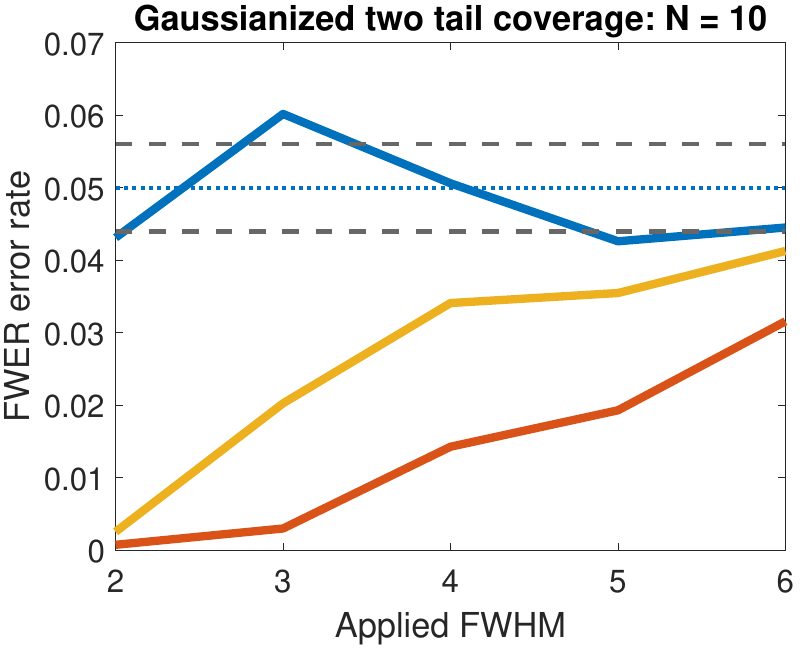}
		\end{subfigure}
		\begin{subfigure}[b]{0.3\textwidth}
			\centering
			\includegraphics[width=\textwidth]{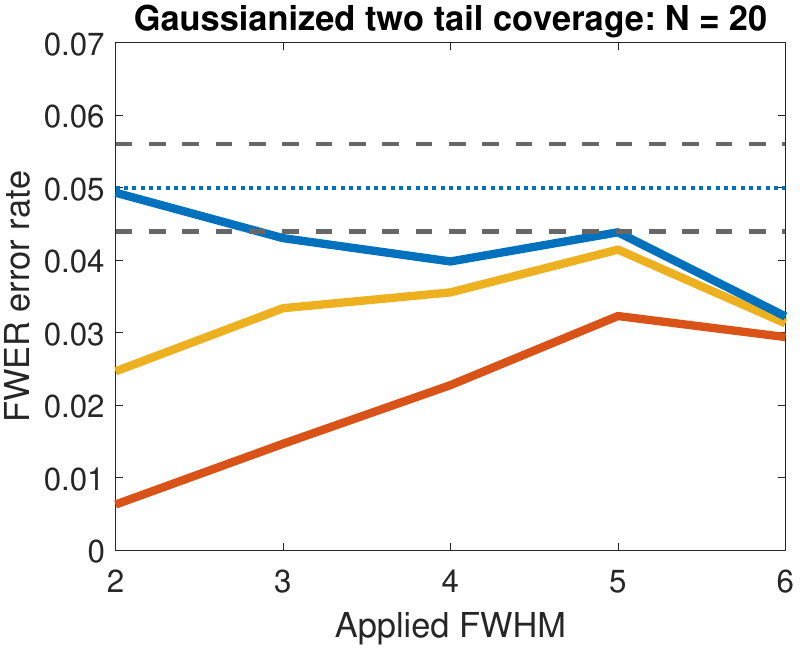}
		\end{subfigure}
		\begin{subfigure}[b]{0.3\textwidth}
			\centering
			\includegraphics[width=\textwidth]{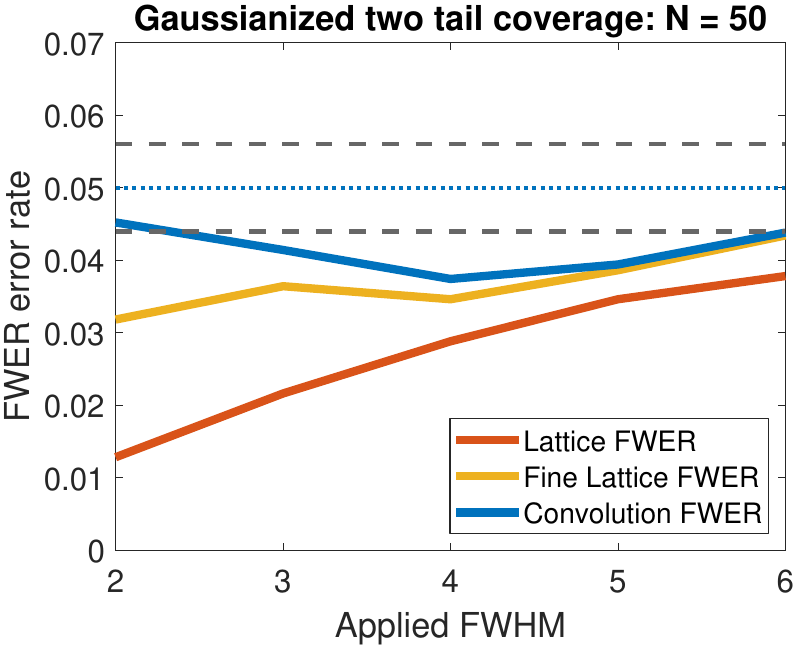}
		\end{subfigure}
	\end{center}
	\caption{Resting state validation of the two tail FWER for $ \alpha = 0.05 $.  For each applied FWHM and $ N \in \left\lbrace 10,20,50 \right\rbrace $ we plot the average error rate that results from applying the methods to 5000 randomly chosen samples of our 7000 resting state subjects. We compare FWER control of the data that is Gaussianized before smoothing (top row) to that of the control for the data that is not Gaussianized (bottom row). The FWER using convolution fields (shown in red) is controlled below the nominal rate in all settings though is more accurate for the Gaussianized data. The ($ r = 0 $) original lattice FWER (shown in purple) is conservative as is, to a lesser extent, the ($ r=1 $) fine lattice FWER (shown in yellow) though this improves as the smoothness increases. The expected number of maxima (shown in blue) above the $ u_{\alpha} $ threshold is accurately predicted for the Gaussianized data.}\label{fig:twotailfwer}
\end{figure}

\begin{figure}[h!]
	\begin{center}
		\begin{subfigure}[b]{0.3\textwidth}
			\centering
			\includegraphics[width=\textwidth]{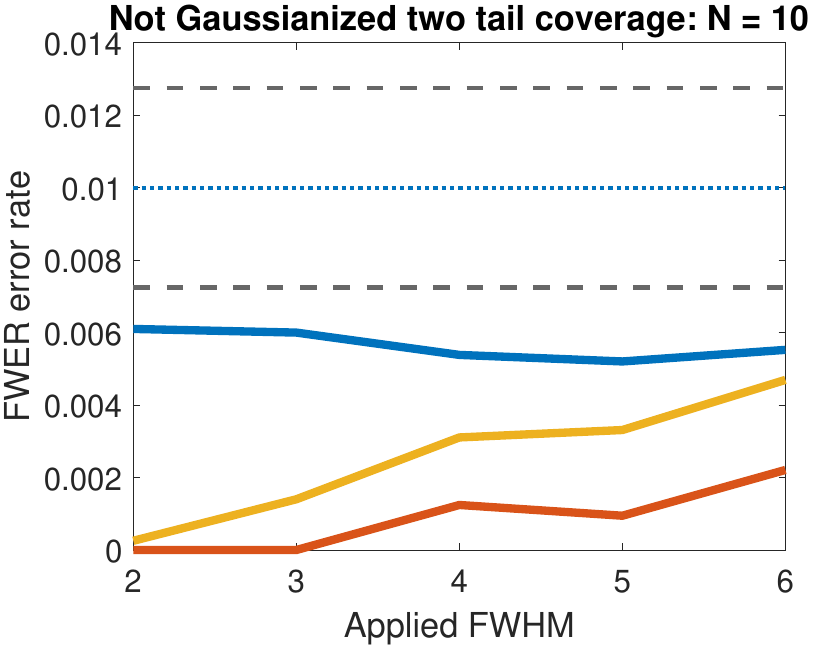}
		\end{subfigure}
		\begin{subfigure}[b]{0.3\textwidth}
			\centering
			\includegraphics[width=\textwidth]{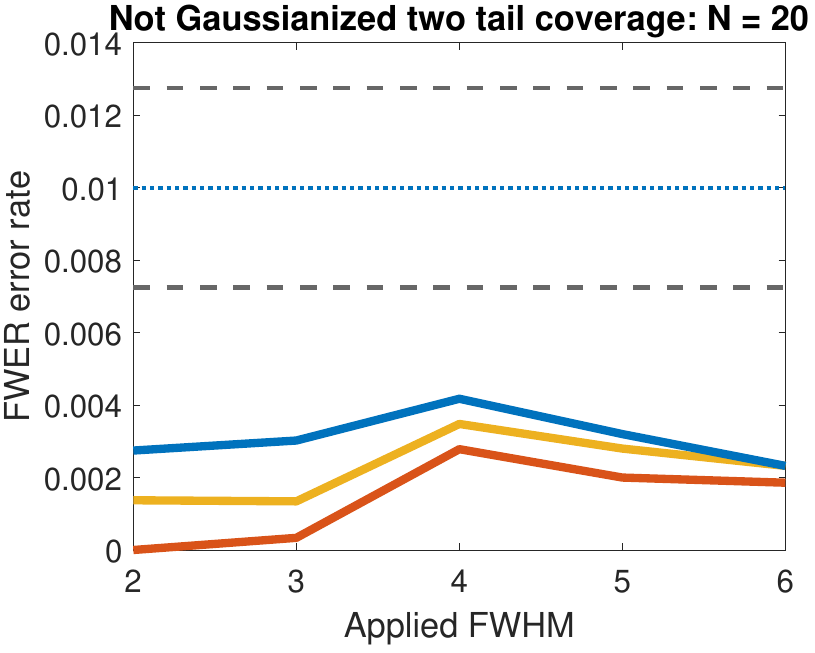}
		\end{subfigure}
		\begin{subfigure}[b]{0.3\textwidth}
			\centering
			\includegraphics[width=\textwidth]{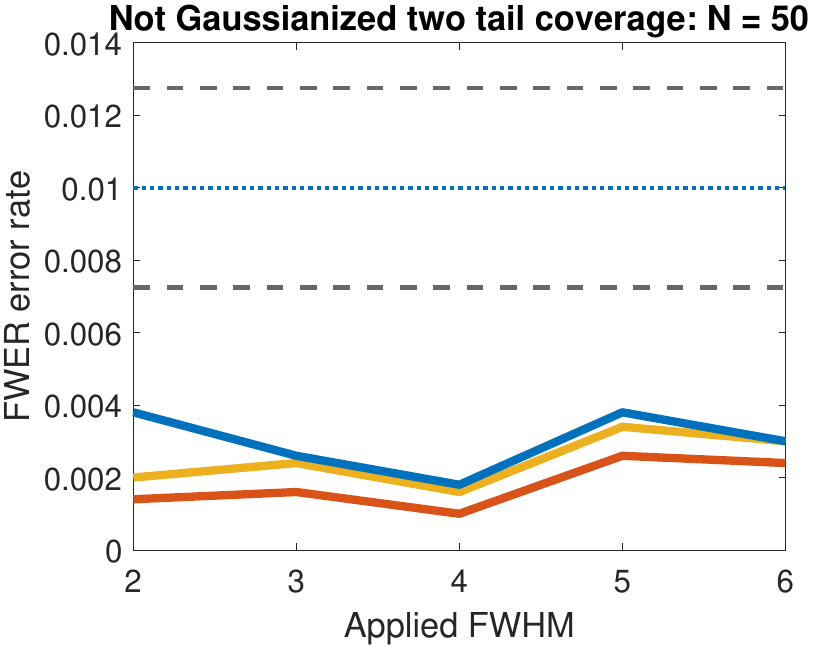}
		\end{subfigure}
	\end{center}
	\vskip\baselineskip
	\begin{center}
		\begin{subfigure}[b]{0.3\textwidth}
			\centering
			\includegraphics[width=\textwidth]{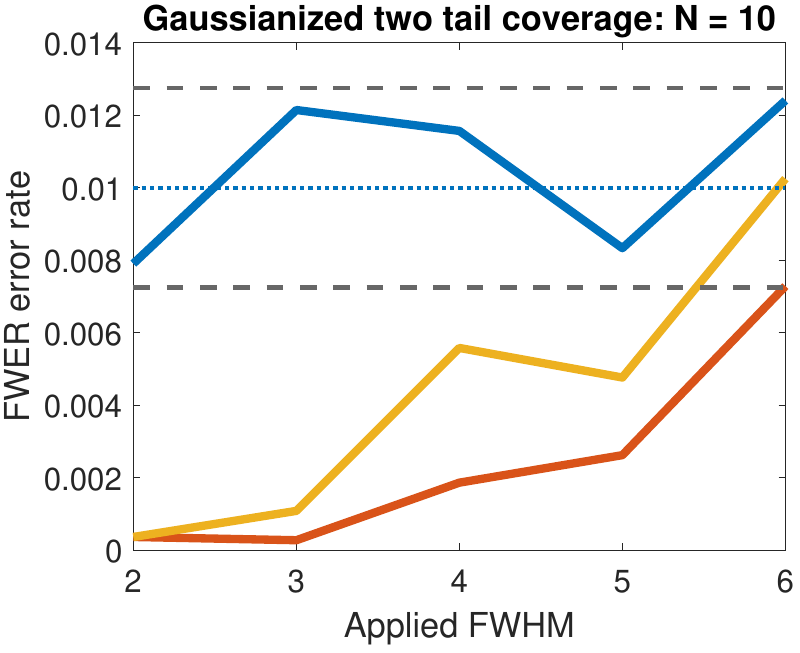}
		\end{subfigure}
		\begin{subfigure}[b]{0.3\textwidth}
			\centering
			\includegraphics[width=\textwidth]{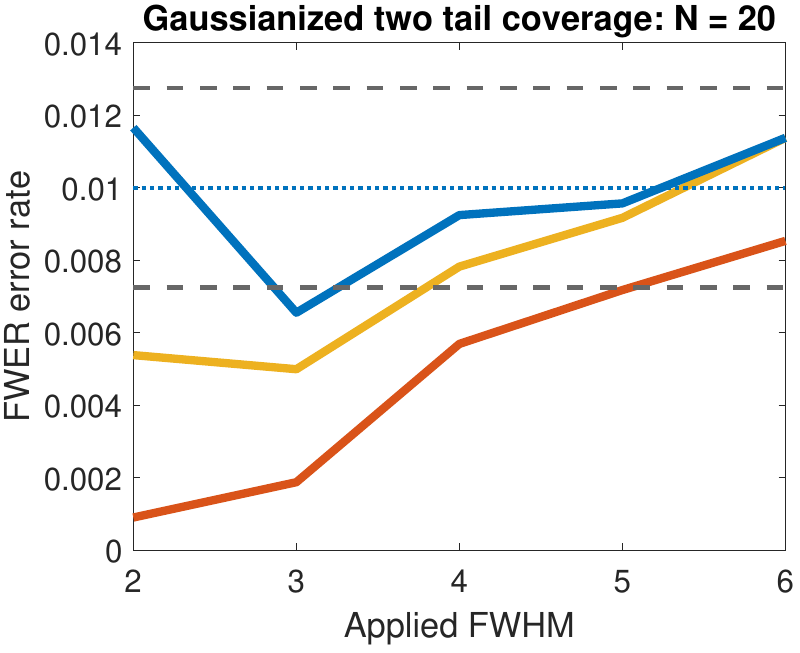}
		\end{subfigure}
		\begin{subfigure}[b]{0.3\textwidth}
			\centering
			\includegraphics[width=\textwidth]{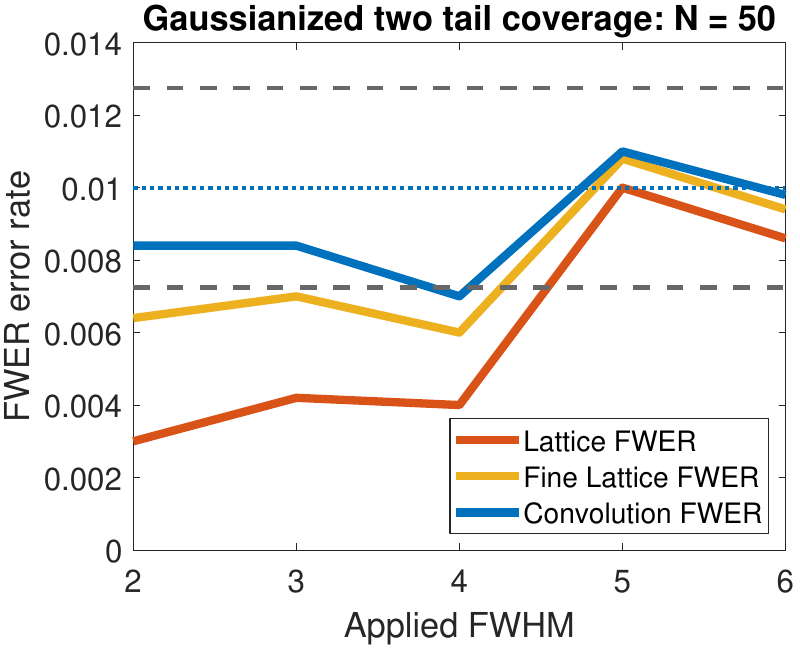}
		\end{subfigure}
	\end{center}	
	\caption{Resting state validation comparing the FWER for $ \alpha = 0.01 $. The settings in each plot are the same as in Figure \ref{fig:twotailfwer}. Controlling at $ \alpha = 0.01 $ requires a higher threshold so the convolution FWER (shown in red) is closer to the nominal level than in Figure \ref{fig:twotailfwer}.}\label{fig:twotailfwer01}
\end{figure}

The results for the FWER validations discussed in Section \ref{rsFWER} are shown for $ \alpha = 0.05 $ and $ \alpha = 0.01 $ in Figures \ref{fig:twotailfwer} and \ref{fig:twotailfwer01} respectively. The first row in the figures illustrates the conservativeness of RFT inference for the original data, even when convolution fields are used. Combining convolution fields with the Gaussianization procedure reduces most of the remaining conservativeness. 

The error rate is controlled to the nominal level in all examples. However the results are still a little conservative in some of the scenarios when $ \alpha  = 0.05$. This occurs because \eqref{eq:onetail} provides an upper bound, see our discussion in Section \ref{S:RFTthresh} for further details. This upper bound becomes an equality at high thresholds because the number of maxima is either 0 or 1 with high probability but it explains the slight discrepancy here. This effect almost disappears when we seek to control the FWER to 0.01: see Figure \ref{fig:twotailfwer01}. This occurs because the thresholds needed to control the FWER to 0.01 are higher and so equation \ref{eq:onetail} becomes a better approximation.

Importantly, the over-coverage that we observed in the simulations for low $ N $ is not present in the real-data validations. We included the simulations to illustrate that in some scenarios some $ N $ is required before convergence. However, the gold standard for inference in fMRI is the resting-state validations. For these the methods work well, in the sense that the EEC is well estimated and false positive rates are controlled below the nominal level, even when $ N $ is as low as $ 10 $.

%
%
%
%
%
%
%
%

%% file: discussion.tex
In this paper we have presented a framework that allows for accurate voxelwise RFT inference under low smoothness and non-Gaussianity, making RFT robust to violations of its traditional assumptions.  Convolution fields allow us to bridge the gap between continuous theory and the lattice data that is collected in practice. Moreover, the Gaussianization procedure allows the GKF to provide accurate inference in low to moderate sample sizes, even when the original data is non-Gaussian. Together these modifications to the standard pipeline allow us to provide a quick and accurate RFT inference framework, for controlling the voxelwise FWER, that is valid in fMRI. 

In this work we validated our methods using what is becoming a gold-standard in the field - deriving noise by regressing fake task designs on resting state data. This approach creates realistic noise distributions, allowing us to understand how methods perform under the high level of non-stationarity and non-Gaussianity that are present in real datasets. Here we have undertaken the largest such study to date involving data from 7000 subjects. Previous resting state validations consisted of substantially smaller sample sizes of at most 198 subjects. As such they suffered from large levels of dependence between draws.

Gaussianization effectively improves the validity of RFT inference if the data is non-Gaussian. Existing RFT theory does not perform well in practice when validated using resting state data. In particular applying it results in a substantial bias for calculations of the expected Euler characteristic.
 This provides a further reason for the conservativeness of traditional RFT voxelwise inference observed in \cite{Eklund2016}, in addition to the explanation provided in Part 1. Instead, comparison of the EC curves after the Gaussianization transformation show a close match (between the expected EC curves and the empirical average EC curves), which is good evidence that the theory is working well for the transformed data. Using the Gaussianized test-statistic random fields for inference thus allows us to reliably control the FWER. 

The substantial level of non-Gaussianity that we found in the resting state fMRI data raises general questions about the validity of other methods that assume Gaussianity of the data or of the test-statistic. These methods include both Bayesian inference which often relies on the assumption that the data is marginally Gaussian (e.g.\cite{Bowman2008}, \cite{Mejia2019}) as well as frequentist inference via RFT (e.g. clustersize inference, \cite{Friston1994}) or resampling methods which often rely on a Central Limit Theorem (e.g. \citep{Winkler2014}). Methods for statistical inference used to analyse fMRI data should, in the light of our resting state data analysis, be tested more rigorously to ensure that they are robust to non-Gaussianity in the data and still provide valid inference. We have shown that the improvements that can result from transforming the data are substantial in the voxelwise RFT inference setting. Thus, we suspect that approaches to address non-Gaussianity may help to improve the validity of other methods which require Gaussianity.

	Our modifications to the traditional RFT framework have the potential to allow other RFT based approaches, such as peak inference \cite{Cheng2015, Schwartzman2019}, to work without requiring high levels of applied smoothness.  Moreover, estimating the EEC correctly is a very important part of cluster size inference \citep{Friston1994}. We have shown here that this can be done in fMRI once the fields have been Gaussianized. Solving cluster failure \citep{Eklund2016} is challenging and beyond of the scope of this work as clustersize inference using RFT requires additional assumptions which may not be reasonable in practice. However this work represents the first, of a number of steps required to make RFT based clustersize inference a reliable method.

%% file: acknowledgements.tex
TEN was supported by the Wellcome Trust, 100309/Z/12/Z and SD was partially funded by the EPSRC. All authors were partially supported by NIH grant R01EB026859. F.T. was funded by the Deutsche Forschungsgemeinschaft (DFG) under Excellence Strategy The
Berlin Mathematics Research Center MATH+ (EXC-2046/1, project ID:390685689). The research was carried out under UK BioBankapplication \#34077, with bulk image data shared within Oxford (with UK BioBank permission) from application 8107.

%% file: appendix.tex
\appendix

\section{Supporting theory for RFT inference}

\subsection{Estimating the LKCs on a voxel manifold}
\subsubsection{Closed forms for  $ \mathcal{L}^G_{D}$ and  $\mathcal{L}^G_{D-1} $}\label{AA:LKCtheory}
Given $ \Lambda $, the covariance matrix of the partial derivatives, closed forms for top two LKCs exist, \cite{Adler2010}. In particular,
\begin{equation}\label{eq:LD}
\mathcal{L}^G_D = \int_S \sqrt{\det\big(\Lambda^G(s)\big)} \, ds
\end{equation}
and
\begin{equation}\label{eq:LD-1}
\mathcal{L}^G_{D-1} = \int_{\partial S} (\det\Lambda^G_{\partial S})^{1/2} \mathcal{H}_{D-1}
\end{equation}
where $ \mathcal{H}_{D-1} $ is the Hausdorff measure of dimension $ D-1 $ on $ \partial S $. Here the matrix $ \Lambda^G_{\partial S} $ corresponds to the covariance of the derivatives of the field $ Y^G_1 $ with respect to the tangent vectors and is defined formally as follows. At each $ t \in \partial S $, let $ e_1(t), \dots,  e_{D-1}(t) \in \mathbb{R}^{D-1}$ be an orthonormal basis to the tangent space to $ S $ at $ t $, then $ \Lambda^G_{\partial S} $ is defined to be the $ (D-1)\times(D-1) $ matrix such for that $ 1 \leq i,j \leq D-1 $,
\begin{equation}\label{boundarylambda}
(\Lambda^G_{\partial S})_{ij}(t) = \cov\left( \frac{\partial a_j(t,x)}{\partial x}\Big|_{x = 0},\frac{\partial a_j(t,x)}{\partial x}\Big|_{x = 0}\right),
\end{equation}	
where $ a_j(t,x) = \frac{Y_1(t + xe_j(t))}{\sigma(t + xe_j(t))}$ for each $ t \in \partial S,$ $ x \in \mathbb{R} $.  Note that changing the ordering of the tangent vectors could affect the definition of $ (\Lambda^G_{\partial S})_{ij}(t) $ as we have defined it. However doing so does not affect $ \det\Lambda^G_{\partial S} $, which the quantity we want to evaluate so everything here is well-defined. 

For the voxel manifold $ S $ in 3D, \eqref{boundarylambda} simplifies substantially as, at almost all\footnote{All except for the set of points that correspond to the $ D-2 $ or lower dimensional subsets of the boundary (i.e. the corners and the edges in 3D). This set is a measure zero subset of $ \partial S $ and so doesn't not contribute to the integral \eqref{eq:LD-1}.} points $ t $ on the boundary, the tangent space is the plane parallel to one of the sides of a voxel on the boundary of the mask. As such, $ \Lambda^G_{\partial S}(t) $ is simply the $ 2\times 2 $ subset of $ \Lambda^G $ corresponding to that plane.

\subsubsection{Estimating the LKCs on the voxel manifold}\label{Lambda2LKC}
Assume that the domain $ S $ is a voxel manifold. Then given an estimate of the smoothness, $ \hat{\Lambda}: S \rightarrow \mathbb{R} $, and a non-negative resolution $ r \in \mathbb{N} $, $ r $ odd, we can estimate $ \mathcal{L}^G_D $ via
\begin{equation*}
\hat{\mathcal{L}}^G_D = \sum_{s \in \mathcal{V}_r} w_r(s)\left| \det(\hat{\Lambda}^G(s)) \right|^{1/2}.
\end{equation*}
Here $ 0 < w_r(s) \leq 1 $ is the volume of $ U_r(s) \cap S $, where $ U_r(s) $ is the $ D $-dimensional cuboid centred at $ s $ with $ d $th side length $ \frac{h_d}{r+1} $ for $ 1\leq d \leq D $. The weights reflect the contribution of each point to the integral. 
When $ D = 3 $ we can estimate $ \mathcal{L}_{D-1} $ as
\begin{equation*}
\hat{\mathcal{L}}^G_{D-1} = \sum_{s \in F^r_{xy}} w'_r(s)\left| \det(\hat{\Lambda}^G_{xy}(s)) \right|^{1/2} + \sum_{s \in F^r_{yz}} w'_r(s)\left| \det(\hat{\Lambda}^G_{yz}(s)) \right|^{1/2} + \sum_{s \in F^r_{zx}} w'_r(s)\left| \det(\hat{\Lambda}^G_{zx}(s)) \right|^{1/2} 
\end{equation*}
where for distinct $ a,b \in \left\lbrace x,y,z \right\rbrace $, $ F^r_{ab} $ denotes the set of points that lie on the faces of the boundary of $ \mathcal{V}_r $ in the $ ab $ plane, $ \hat{\Lambda}_{ab} $ denotes the $ 2 \times 2 $ matrix $ ab $ subset of $ \hat{\Lambda} $ and $ w'_r(s) $ is the area of the intersection of $ U_r(s) \cap \partial S $ with the $ ab $ plane. Traditional RFT methods take $ r = 0 $ as they do not use convolution fields. Under stationarity and taking $ r = 0 $, the estimates of the top two LKCs reduce to
\begin{equation*}
\hat{\mathcal{L}}^G_D = \left| \det(\hat{\Lambda}^G) \right|^{1/2}\left| \mathcal{V} \right|
\end{equation*}
and 
\begin{equation*}
\hat{\mathcal{L}}^G_{D-1} = \left| F_{xy} \right|\left|\det(\hat{\Lambda}^G_{xy}) \right|^{1/2} + \left| F_{yz} \right|\left| \det(\hat{\Lambda}^G_{yz}) \right|^{1/2} + \left| F_{zx} \right|\left| \det(\hat{\Lambda}^G_{zx}) \right|^{1/2}.
\end{equation*}
where for distinct $ a,b \in \left\lbrace x,y,z \right\rbrace $, $ \left| F_{ab} \right| $ is the volume of the $ ab $ faces on the boundary of $ \mathcal{V}_r $.
%
\subsection{The Kiebel estimate of $ \Lambda $ and the LKCs}\label{SS:remmet}
\cite{Worsley1992} and \cite{Kiebel1999} assumed that the only values available to use for inference on the smoothness were i.e. $ (Y_n(v))_{v \in \mathcal{V}, 1\leq n \leq N} $. These correspond to the original convolution fields evaluated on the lattice in our notation for a given number of subjects $ N \in \mathbb{N} $. As such their estimates of $ \Lambda $, the covariance matrix of the original fields, use discrete derivatives. In particular, on the lattice the partial derivatives of the residuals in the $ d $th direction can be estimated via 
\begin{equation}\label{eq:DD}
Z_{n,d}(v) = \left( R_n(v+h_d\delta_d) - R_n(v) \right)/h_d
\end{equation}
where $ \delta_d$ is the unit vector in the $ d $th direction and $ h_d $ is the difference between lattice points in the $ d $th direction. Here $ R_n = \left(Y_n - \frac{1}{N}\sum_{n = 1}^N Y_n\right)\left(\frac{1}{N-1}\sum_{n = 1}^N \left( Y_n - \frac{1}{N}\sum_{n = 1}^N Y_n \right)^2\right)^{-1/2}$ are the residual fields based on the original convolution fields. Then, under stationarity, the elements of $ \Lambda $ can be estimated by taking (for $i,j = 1, \dots, D$),
\begin{equation}\label{eq:latsmooth}
(\hat{\Lambda}_K)_{ij} = \frac{N-3}{(N-2)(N-1)\left| \mathcal{V} \right|}\sum_{v \in \mathcal{V}}\sum_{n = 1}^N Z_{n,i}(v) Z_{n,j}(v) 
\end{equation}
where the inner sum is taken over all  $ v $ such that $ v, v+h_i\delta_i, v+h_j\delta_j \in \mathcal{V} $ and $ \left| \mathcal{V} \right| $ denotes the number of voxels in $ \mathcal{V} $. Compare \eqref{eq:latsmooth} to equation (14) of \cite{Kiebel1999}. We refer to this estimate in the main text as the \textbf{Kiebel estimate} of $ \Lambda $. Note that we have scaled by $N-1$ instead of $ N $ to account for the fact that we have subtracted the mean in \eqref{eq:resid}.


The estimate $ \hat{\Lambda}_K $ can be used to estimate the LKCs on the voxel manifold, by taking $ \hat{\Lambda}(s) = \hat{\Lambda}_K$ for all $ s \in S $ and plugging this into the formulae for the LKCs, see the stationary forms at the end of Section \ref{Lambda2LKC}. We refer to the resulting estimates as the Kiebel estimates of the LKCs. These estimates are plugged into the GKF and used to provide the Kiebel estimate of the EEC for the original data: this is the estimate used in the first row of plots in Figures \ref{fig:EECplotsFWHM5} and \ref{fig:EECplotsFWHM2}. Using $ R_n^G $ instead of $ R_n $ in \eqref{eq:DD} and \eqref{eq:latsmooth} provides a Gaussianized version of the Kiebel estimate of $ \Lambda $ and corresponding estimates of the LKCs. These estimates are similarly used to provide of the EEC for the Gaussianized data: this is the estimate used in the second row of plots in Figures \ref{fig:EECplotsFWHM5} and \ref{fig:EECplotsFWHM2}. As can be seen from these figures, these estimates are not reliable in fMRI. 

The other method typically used for calculating the LKCs under stationarity is the Forman estimate. This depends directly on the Forman estimate of the FWHM and so we defer discussion of this until Section \ref{SS:FormanFWHM}.

\begin{remark}
		The assumption of stationarity is crucial for the Kiebel estimate because in \eqref{eq:latsmooth} we averaged the covariance estimates over the whole lattice $ \mathcal{V}. $ This assumption is problematic in fMRI and so using $ \hat{\Lambda}_K $ to estimate the LKCs and thus the expected Euler characteristic leads to substantial bias (as we show in Figures \ref{fig:EECplotsFWHM5} and \ref{fig:EECplotsFWHM2}).
\end{remark}

\begin{remark}\label{rmk:correctionfactor}
	In order to obtain an unbiased estimate of $ \Lambda $ from the residual fields we multiplied by the correction factor $ \frac{N-3}{N-2} $ in \eqref{eq:latsmooth}. This factor is required in order to account for the fact that we divide by $ \hat\sigma_N $ rather than $ \sigma $ in \eqref{eq:resid}), see e.g. \cite{Worsley1996c} and distinguishes the Kiebel estimate of $ \Lambda $ from the estimates of \cite{Worsley1992} which do not include the correction factor. The Kiebel estimate of $ \Lambda $ averages across the whole domain and so, even under stationarity, if the correction factor is not included it will result in a biased estimate of the LKCs. 
	
	Importantly we didn't need to include the correction factor in \eqref{eq:lambdahatconv} in order to obtain the estimate of $ \Lambda $ that we used to calculate the LKCs. This is because in \eqref{eq:lambdahatconv} $ \Lambda $ is not averaged over the domain and so the estimate does not pick up bias in the same way. In fact using \eqref{eq:lambdahatconv}, produces an unbiased estimate of the LKCs even though it is a biased estimator for $ \Lambda $ itself, see \cite{Taylor2007} and Theorem 3 of \cite{Telschow2023FWER}.
\end{remark}

\subsection{Interpretation and accuracy of the EEC approximation}
\subsubsection{One-sided inference}\label{A:oneside}
When performing one-sided inference, we are seeking to identify regions of $ S $ for which $ \mu_G > 0 $. In this case we define our null set as $ S_0 = \lbrace s \in S: \mu_G(s) \leq 0 \rbrace$ be the subset of $ S $ on which $ \mu $ is zero. Then we define the familywise error rate (FWER) of this rejection procedure to be
\begin{equation*}
\text{FWER} = \mathbb{P}\left(\sup_{s \in S_0} |T^G(s)| > u\right)
\end{equation*}
to be the probability of a single super-threshold excursion. We can then control the FWER using RFT in a similar way to how we did in Section \ref{S:RFTthresh}, because 
\begin{equation}\label{eq:onetail}
\mathbb{P}\left( \sup_{s \in S} T^G(s) >  u_{2\alpha} \,|\,S = S_0\right) \leq \mathbb{E}\left[ M_{u_{2\alpha}}(T^G)\,|\, S = S_0 \right] \leq  \mathbb{E}\left[ \chi(\mathcal{A}_{u_{2\alpha}}(T^G))\,|\, S = S_0 \right] \approx \alpha,
\end{equation}
where the threshold $ u_{2\alpha} $ is defined via equation \eqref{eq:LKCu}.

\subsubsection{Understanding the accuracy of the EEC approximation}\label{AA:accuracy}
If $u$ is large enough then each connected component of $ \mathcal{A}_{u}(T^G) $ contains a single maximum and no other critical point of $T$. Each critical point of $ T^G $ within $ \mathcal{A}_{u}(T^G) $ affects the EC of the excursion set and so at high thresholds $ u, $ $M_{u}(T^G) = \chi(\mathcal{A}_{u}(T^G))$ because the connected components each contain a single local maxima. As such at the thresholds
used in FWER inference we expect the approximation, 
$ \mathbb{E}\left[ M_{u_{\alpha}}(T^G) \right] \approx \mathbb{E}\left[ \chi(\mathcal{A}_{u_{\alpha}}(T^G)) \right] $
to be very accurate.

As shown in Section \ref{S:RFTthresh}, $ 2\mathbb{E}\left[ M_{u_{\alpha}}(T^G)\,|\,S_0 = S \right] $ provides an upper bound on the probability $ \mathbb{P}\left( \sup_{s \in S} \left| T^G(s) \right| >  u_{\alpha/2} \,\middle|\, S_0 = S\right)$. However, for the thresholds used for FWER inference we expect to have 0 or 1 maxima above the threshold with high probability, since we are controlling the expected number of maxima to a level $ \alpha \leq 0.05. $ As such it is in fact reasonable to make the approximation 
\begin{equation*}
	\mathbb{P}\left( \sup_{s \in S} T^G(s) \geq u \,\middle|\, S_0 = S\right) \approx \mathbb{E}\left[ M_{u_{\alpha}}(T^G)|S_0 = S \right].
\end{equation*}
Moreover, at these thresholds the probability 
\begin{equation*}
	 \mathbb{P}\left(\sup_{s \in S} T^G(s) \geq u, \min_{s \in S} T^G(s) < -u\right) 
\end{equation*}
is very small. We thus expect the upper bound provided by $ \mathbb{E}\left[ M_{u_{\alpha}}(T^G)\,|\,S_0 = S \right] $ to be relatively tight. The accuracy of the approximation improves as $ \alpha $ decreases since the FWER threshold $ u_\alpha $ increases. More precise asymptotics justifying these approximations can be found in \cite{Taylor2006}.

\section{The good lattice assumption}
\subsection{Traditional voxelwise RFT Inference}\label{SS:tradRFT}
Traditional RFT works with fields and a test-statistic defined on the lattice $ \mathcal{V} $ instead of an open set $ S $. For this section this $ T_L $ be this test-statistic evaluated on the lattice. The traditional RFT voxelwise framework infers on the probability that the test-statistic $ T_L $ exceeds a threshold $ u \in \mathbb{R} $. In particular, by Markov's inequality,
\begin{equation*}
\mathbb{P}\left( \max_{v \in \mathcal{V}} T_L(v) \geq u  \right) = \mathbb{P}\left( M_u(T_L) \geq 1 \right)\leq \mathbb{E}\left[ M_u(T_L) \right].
\end{equation*}
Here $ M_u(T_L) $ is the number of local maxima of $ T_L $ which are greater than or equal to $ u $. If $ u $ is chosen such that the expectation equals $ \alpha $ then the probability that the test-statistic exceeds $ u $ will be less than or equal to $ \alpha. $ At high thresholds $ u $ the number of local maxima above the threshold is 0 or 1 with high probability and so
\begin{equation*}
\mathbb{P}\left( M_u(T_L) \geq 1 \right) \approx \mathbb{E}\left[ M_u(T_L) \right].
\end{equation*}
However, even at low thresholds the expected number of maxima still serves as an upper bound.

\subsection{Good lattice assumption}
There are no direct formulae available for the expected number of local maxima of a test-statistic on a lattice. However these quantities are well studied for random fields defined on a continuous domain \citep{Adler1981, Adler2007}. In particular for random fields it is possible to closely approximate the expected number of local maxima above a level $ u $ using the EEC of the excursion set which has a simple closed form known as the GKF \citep{Taylor2006}, see Section \ref{SS:EEC}.

As such in order to provide a bound on $ \mathbb{P}\left( \max_{v \in \mathcal{V}} T_L(v) \geq u  \right) $ traditional RFT inference implicitly makes the following key assumption.\\

\textbf{Good Lattice Assumption:} There is a continuous set $ S \supset \mathcal{V}$ such that  $ \lbrace T_L(v): v \in \mathcal{V} \rbrace$ extends to a random field $ \lbrace T(s): s \in S \rbrace$ such that $ T(v) = T_L(v) $ for all $ v \in \mathcal{V} $. In particular on this set
\begin{equation*}
\max_{v \in \mathcal{V}} T(v) \approx \sup_{s \in S} T(s)
\end{equation*}
and letting $ M_u(T) $ be the number of local maxima of the continuous field $ T $ above $ u $, 
\begin{equation}\label{ass:gl}
\mathbb{E}\left[ M_u(T_L) \right] \approx \mathbb{E}\left[ M_u(T) \right].
\end{equation}
The main problem with this assumption is that it is not precise in terms of how good the approximation must be. However, when the data is sufficiently smooth this assumption does not seem unreasonable, which is why RFT inference has historically always required a high level of smoothness. Even though the extension $ T $ has never been explicitly defined in the literature traditional RFT implicitly provides estimates of $ \mathbb{E}\left[ M_u(T) \right] $ at high thresholds $ u $ through the use of the GKF and thus provides a bound on $ \mathbb{P}\left( \max_{v \in \mathcal{V}} T_L(v) \geq u  \right) $.

\subsection{The conservativeness of traditional inference}\label{SS:conserve}
For any extension $ T$ the maximum of the test-statistic on the lattice is bounded by the maximum on $ S $, i.e
\begin{equation}\label{eq:supbound}
\max_{v \in \mathcal{V}} T(v) \leq \sup_{s \in S} T(s)
\end{equation}
At the smoothing bandwidths typically used in fMRI it is not reasonable to assume approximate equality in \ref{eq:supbound} and so the good lattice assumption breaks down. Given $ \alpha \in (0,1) $, traditional RFT inference chooses $ u $ to control $ \mathbb{P}\left( \sup_{s \in S} T(s) \geq u \right) $ at the level $ \alpha $. As such the threshold needed to control excursions of the continuous process $ T $ is higher than that needed for $ T_L $. This leads to conservative inference and a corresponding loss of power since \eqref{eq:supbound} implies that
\begin{equation*}
\mathbb{P}\left( \max_{v \in \mathcal{V}} T_L(v) \geq u \right)
=
\mathbb{P}\left( \max_{v \in \mathcal{V}} T(v) \geq u \right)	
< \mathbb{P}\left( \sup_{s \in S} T(s) \geq u \right).
\end{equation*} 
This problem with traditional RFT is well known and it has been shown that a high level of applied smoothing is required to avoid excessive conservativeness (\cite{Hayasaka2003} and \cite{Worsley2005}). Eklund tests of voxelwise RFT demonstrate that this effect can be quite pronounced, see Section \ref{SS:fwer} and \cite{Eklund2016}'s Figure 1 (rightmost panels). 



\section{Estimating the FWHM}\label{A:latsmo}
\subsection{Understanding the background of FWHM/LKC estimation in neuroimaging}
Historically the neuroimaging community has focused on estimating the FWHM \citep{Jenkinson2000}, using this to calculate an estimate $ \hat{\Lambda} $ of $ \Lambda $ and plugging in $ \hat{\Lambda} $ to obtain an estimate of the LKCs. This approach uses a specific relationship between $ \Lambda $ and the FWHM which holds when the noise field arises from smoothing white noise with a Gaussian kernel, see Section \ref{SS:FWHMest}. This is not a good model of fMRI data as the noise in fMRI is non-stationary \cite{Eklund2016}. As such it makes more sense to target $ \Lambda $ directly without the need to estimate the FWHM. Even if the field were stationary, the focus on estimating the FWHM is misguided because it is the LKCs not the FWHM that are relevant for performing inference using RFT. Direct estimates of $\Lambda$ should be plugged into the formulas for the LKCs, rather than estimating $ \Lambda $ by first estimating the FWHM. 

There are two main methods in the fMRI literature for estimating $ \Lambda $: those of \cite{Kiebel1999} and \cite{Forman1995}, defined in Sections \ref{SS:remmet} and \ref{SS:FormanFWHM} respectively. Both of these give biased estimates of $ \Lambda $ and the EEC when applied to fMRI data. We discuss these approaches and their assumptions in what follows and describe how they have historically been used to estimate the LKCs and the FWHM. We also introduce a new estimate of the FWHM itself that uses convolution fields.

\subsection{White noise model}
Before \cite{Taylor2006} came up with the GKF for non-stationary random fields and it was applied in neuroimaging \cite{Taylor2007}, a number of papers on RFT based inference assumed that the noise has the same distribution as Gaussian white noise convolved with a kernel \cite{Worsley1992, Friston1994, Kiebel1999, Worsley2005}. Formally this distribution arises as follows.
\begin{definition}
	(Convolved Gaussian white noise model) Let $ W:\mathbb{R}^D \rightarrow \mathbb{R}$ be a $D$-dimensional Weiner random field. Let $ K: \mathbb{R}^D \rightarrow \mathbb{R} $ be a $ D $-dimensional kernel. Then we can define a convolved white noise random field as $ Y: \mathbb{R}^D \rightarrow \mathbb{R} $ such that
	\begin{equation}\label{mod:cwnm}
	Y(s) = \int_{\mathbb{R}^D} K(s,t) dW_t
	\end{equation}	
	where the integral is in the Ito sense.
\end{definition}
\cite{Holmes1994} showed that additional more restrictive assumptions on the field can be used to derive a closed form for $ \Lambda $, i.e.
\begin{theorem}\label{thm:asd}
	In this setting of the convolved Gaussian white noise model and that $ K $ is $ D $-dimensional Gaussian kernel with covariance $ \Sigma. $ Then $$\Lambda = {\rm cov}(\nabla Y(s)^T)  = \Sigma^{-1}/2, \text{ for all } s \in \mathbb{R}^D.$$
	In particular if $ \Sigma $ is diagonal, then 
	\begin{equation*}
	\Lambda = \pmatrix{\Sigma_{11}^{-1}/2. 0. 0; 0. \Sigma_{22}^{-1}/2. 0; 0. 0. \Sigma_{33}^{-1}/2} = \pmatrix{1/{\rm FWHM}_1^2. 0. 0; 0. 1/{\rm FWHM}_2^2. 0; 0. 0. 1/{\rm FWHM}_3^2}4\log(2) 
	\end{equation*}
	where {\rm FWHM}$_d $ is the full width have maximum of $ K $ in the $ d $th direction.
\end{theorem}
Historically the relationship between $ \Lambda $ and the FWHM under this model has led to an emphasis on FWHM estimation when performing RFT inference in
fMRI \citep{Worsley1992, Worsley1996, Kiebel1999}. Assume that data from this model is observed on a set $ S $ with volume $ |S| $ and surface area $ |\partial S | $ and that the associated kernel is Gaussian and isotropic. Then given an estimate $ f $ of the FWHM (which is the same in each direction in this case), estimates of the LKCs can for instance be obtained as
\begin{equation}\label{eq:LKCFWHM}
	\hat{\mathcal{L}}_D = 4^{\frac{D}{2}}\log(2)^{\frac{D}{2}}f^D|S| \quad \text{ and } \hat{\mathcal{L}}_{D-1} = 4^{\frac{D-1}{2}}\log(2)^{\frac{D-1}{2}}f^{D-1}|\partial S|.
\end{equation}
This general approach - to first estimate the FWHM and then use this to estimate the LKCs - is the standard way of obtaining the LKCs in both SPM and FSL. In our view it is misguided because for RFT the LKCs are the objects of interest not the FWHM and an unbiased estimate of the FWHM in fact provides a biased estimate of $\Lambda$ and the LKCs \citep{Telschow2023FWER}. Moreover the relationship between $ \Lambda $ and the FWHM only holds under the restrictive model described in Theorem \ref{thm:asd} which is not a realistic model of the noise in fMRI \citep{Eklund2016}.

The full LKC estimation framework that we describe in Section \ref{SS:LKCest} is valid under any type of non-stationarity and so is thus much more general and widely applicable than focusing on estimating the LKCs using the FWHM. Using this framework, when combined with our Gaussianization procedure, provides more reliable estimates of the LKCs (as shown for instance in Figures \ref{fig:EECplotsFWHM5} and \ref{fig:EECplotsFWHM2}). Thus in the context of performing RFT inference we recommend targetting the LKCs directly rather than first estimating the FWHM and then using that estimate to calculate the LKCs.

In the following sections we describe the Kiebel and Forman methods for estimating the FWHM and how to derive the corresponding LKC estimates. In the case that the FWHM is in fact the object of interest (rather than the LKCs) we show that convolution fields can be used to provide an improved estimate of the FWHM. We compare the performance of the Kiebel, Forman, and convolution estimators for the FWHM in Section \ref{SS:smoest} using a numerical simulation. 



\subsection{The Kiebel estimate of the FWHM}\label{SS:FWHMest}
In the setting of Theorem \ref{thm:asd}, given an estimate $ \hat{\Lambda} $ of $ \Lambda $, the FWHM in the $ d $th direction can be estimated as $\widehat{\text{FWHM}}_d = \hat{\Lambda}_{dd}^{-1/2}\sqrt{4\log(2)}$. If the Kernel is further assumed to be isotropic (i.e. $ \Sigma $ is a multiple of the identity) then the FWHM is the same in each direction and can therefore be estimated via
\begin{equation}\label{eq:FWHM}
\widehat{\text{FWHM}} = \left( \frac{4\log(2)}{\frac{1}{D}\sum_{d = 1}^D\hat{\Lambda}_{dd}} \right)^{1/2}.
\end{equation} Plugging in the lattice estimate (\ref{eq:latsmooth}) for $ \hat\Lambda $ yields the Kiebel estimate of the FWHM.
 
\subsection{The Forman estimates of the FWHM and the LKCs}\label{SS:FormanFWHM}
The Kiebel estimate of the FWHM discussed above relies on discrete derivatives on the lattice to estimate $ \Lambda $, as described in \ref{eq:latsmooth}. The Forman approach instead assumes that the model described in Theorem \ref{thm:asd}, but that only $ \lbrace Y(v): v \in \mathcal{V}\rbrace $ is observed rather than the full field $ Y $.

When the smoothing kernel is Gaussian and isotropic the  resulting Forman estimate of the FWHM is given by
\begin{equation*}
\widehat{\text{FWHM}}=\left( \frac{-2\log(2)}{\log\left( 1- \frac{1}{2D}\sum_{d= 1}^D (\hat{\Lambda}_K)_{dd} \right)} \right)^{\frac{1}{2}}
\end{equation*}
where $ \hat{\Lambda}_K $ is the Kiebel estimate of $ \Lambda $. Note that this estimate, by definition, includes the scaling factor of $ \frac{N-3}{N-2} $. The original Forman estimator appears, in fact, not to have accounted for the scaling factor: we include the factor here because it leads to a better estimate of the FWHM (see Figure \ref{fig:FWHMest}). Using this estimate for the FWHM we can in turn obtain an alternative estimate of $ \Lambda $ defined as the $ D \times D $ matrix $ \Lambda_F $ such that $ (\hat\Lambda_F)_{ij} = 4\log(2)/\widehat{\text{FWHM}} 1_{i = j} $ for $ 1 \leq i,j \leq D. $ Estimates of the LKCs can then be obtained by plugging $ \widehat{\text{FWHM}} $ into the formulae given in \eqref{eq:LKCFWHM}.

When $ \Sigma $ is diagonal but non-isotropic, the smoothness in the $ d $th direction can be estimated via $\widehat{\text{FWHM}}_d = \sqrt{2\log(2)}\left( -\log\left( 1- \frac{1}{2}\hat{\Lambda}_{dd} \right) \right)^{-\frac{1}{2}}$. See \cite{Jenkinson2000} for a detailed derivation of this estimator. Correspondingly it is possible to derive a non-isotropic Forman estimate of $ \Lambda, $ and correspondingly estimates of the LKCs.
\subsection{The convolution estimate of the FWHM}\label{convFWHM}
The convolution framework can in principle be used to estimate the FWHM under the model described in Theorem \ref{thm:asd}, rather than the LKCs, if this is the object of interest. In this case we can define $  \hat{\Lambda}': S \rightarrow \mathbb{R}^D \times \mathbb{R}^D$ defined via
\begin{equation}\label{eq:CL}
\hat{\Lambda}'_{ij}(t) = \frac{N-3}{(N-2)(N-1)}\sum_{n = 1}^N \frac{\partial R_n(t)}{\partial t_i}\left( \frac{\partial R_n(t)}{\partial t_j} - \frac{1}{N}\sum_{m = 1}^N\frac{\partial R_m}{\partial t_j} \right)
\end{equation}
for $ i,j = 1, \dots, D, $ $ t \in S. $ Then under stationarity, we can compute 
\begin{equation*}
 	\hat{\Lambda}_C = \sum_{v \in \mathcal{V}}\hat{\Lambda}'_{ij}(v)
\end{equation*}
in order to estimate $ \Lambda $. Here we included the correction factor discussed in Remark \ref{rmk:correctionfactor} because the average is taken over the whole domain. $ \hat{\Lambda}_C $ can be used to calculate the FWHM as in Section \ref{SS:FWHMest} to provide a convolution estimate of the FWHM. As we show in Section \ref{SS:smoest} this provides an unbiased estimate even at low levels of applied smoothness. 

When it comes to the LKCs, under non-stationarity $ \hat{\Lambda}'(t) $ provides an unbiased estimate of $ \Lambda(t) $ for each $ t \in S. $  This can be seen by arguing as in \citep{Worsley1996c}. However using $ \hat{\Lambda}' $ in Section \ref{Lambda2LKC} instead of $ \hat{\Lambda} $ would result in a biased estimate of the LKCs even under stationarity. This follows e.g. from \cite{Telschow2023FWER}'s Theorem 3 which shows that using $\hat{\Lambda} $ provides unbiased estimates of the LKCs.

\subsection{Comparing FWHM estimation methods}\label{SS:smoest}
\begin{figure}[h!]
	\begin{center}
		\begin{subfigure}[b]{0.4\textwidth}
			\centering
			\includegraphics[width=\textwidth]{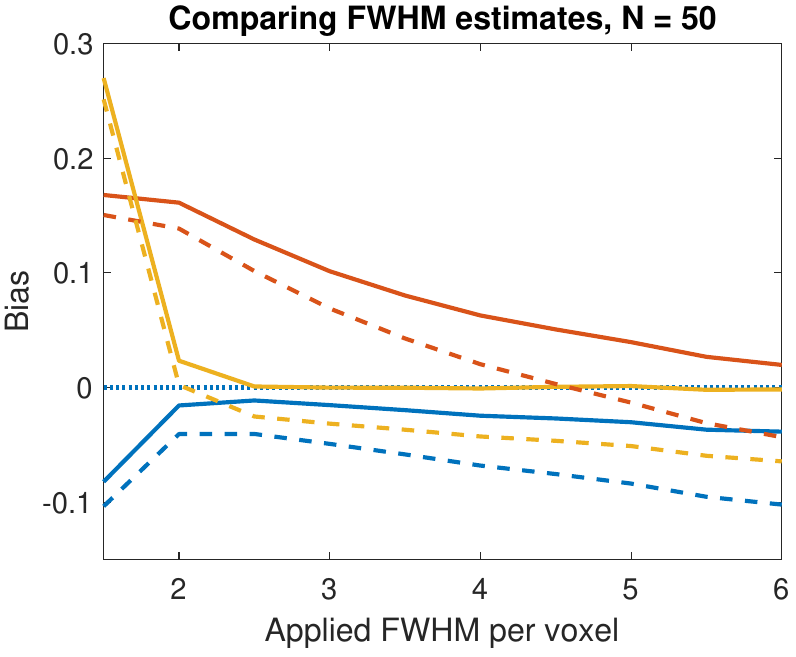} 
		\end{subfigure}
		\hspace{1.5cm}
		\begin{subfigure}[b]{0.4\textwidth}  
			\centering 
			\includegraphics[width=\textwidth]{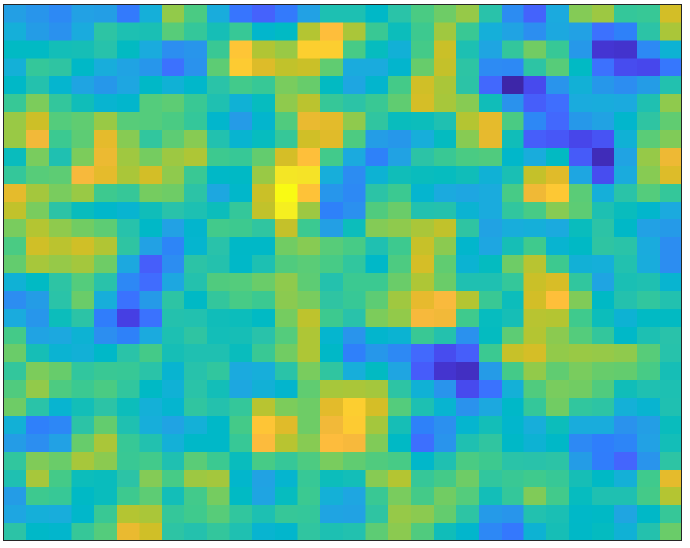} 
		\end{subfigure}
	\end{center}
	\vskip\baselineskip
	\begin{center}
		\begin{subfigure}[b]{0.4\textwidth}
			\centering
			\includegraphics[width=\textwidth]{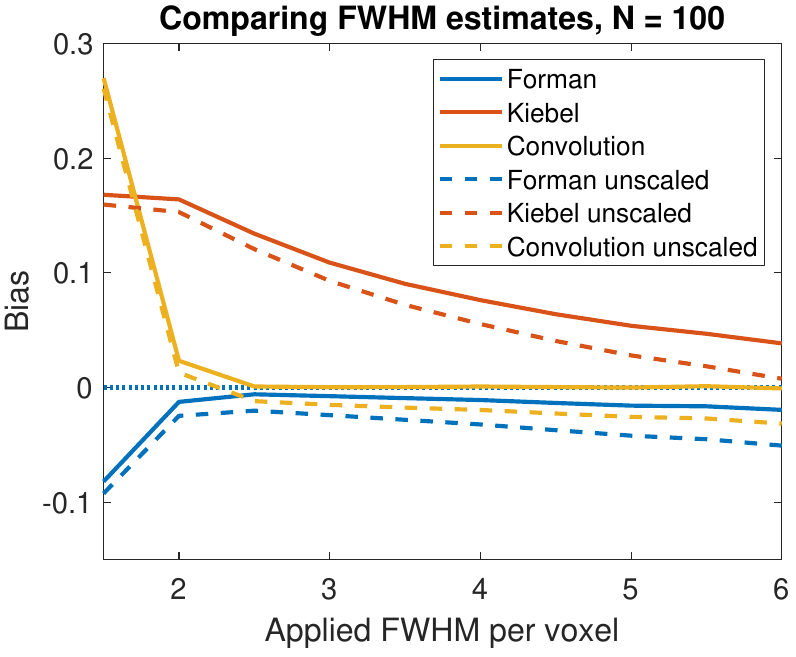} 
		\end{subfigure}
		\hspace{1.5cm}
		\begin{subfigure}[b]{0.4\textwidth}  
			\centering 
			\includegraphics[width=\textwidth]{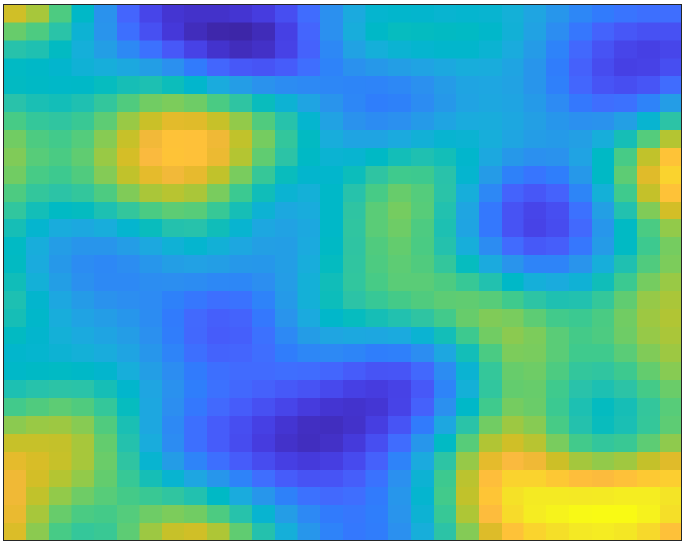} 
		\end{subfigure}
	\end{center}
	\caption{Comparing FWHM estimation for Gaussian white noise smoothed with a diagonal isotropic Gaussian kernel. On the left we plot the FWHM estimation bias (estimated-applied) against the applied FWHM for the 3 different methods (obtained using $ N = $ 50 (top left) and 100 (bottom left) subjects).  The convolution estimate is essentially unbiased when the applied FWHM is greater than or equal to $ 2.5 $. The Forman and Kiebel estimates are both biased at all FWHM though the bias of the Kiebel estimator decreases as the smoothness increases. These graphs illustrate the importance of the $ \frac{N-3}{N-2} $  scaling factor: the unscaled results (dashed lines) omit this factor and result in a downward bias. On the right we plot two 2D slices through 3D white noise smoothed with FWHM = 2 voxels (top right) and 6 voxels (bottom right) on original ($ r = 0 $) lattice.}\label{fig:FWHMest}
\end{figure}

In this section we compare the different methods discussed above for estimating the FWHM. In order to compare their performance we generate random fields on a $ 30 \times30 \times 30 $ lattice by smoothing i.i.d Gaussian white noise with a diagonal isotropic Gaussian kernel. For each applied FWHM in $ \left\lbrace 2, 2.5, \dots, 6 \right\rbrace $ we generate $ N = 50$ and $ 100 $ random fields to yield an estimate of the smoothness. We do this 1000 times and take the average over the estimates for each $ N $ and each FWHM. The results are shown in Figure \ref{fig:FWHMest}. Note that to correct for the edge effect we simulate data on a larger lattice (increased by a size of at least 4 times the standard deviation of the kernel in each direction) and take the central lattice subset as discussed in \cite{Davenport2022Ravi}.

The results in Figure \ref{fig:FWHMest} show that the Kiebel estimates are positively biased and the Forman estimates are negatively biased, though the bias of the Kiebel estimator decreases as the smoothness increases. The convolution estimates have negligible bias except for low FWHM. The reason for the bias at low applied smoothness is that the convolution fields generated are technically non-stationary. Low smoothness is not a problem for LKC estimation. However the FWHM is only well defined for stationary random fields and so we wouldn't expect a meaningful estimate in this case. For larger FWHM the convolution fields are still technically non-stationary however in practice they are very close to being stationary \citep{Telschow2023FWER} so the FWHM can be accurately estimated.

We have also plotted the estimates of the FWHM that result when the estimate for $ \Lambda $ is not scaled by $ \frac{N-3}{N-2} $.  Asymptotically the scaling factor is irrelevant but it has a notable effect at low $ N $. For the Kiebel estimate, at these smoothness levels, not scaling leads to an artificial improvement but causes bias when the applied smoothing is higher. The fact that the convolution estimates appear unbiased when the scaling factor is applied provides a clear indication that scaling by the factor is the correct approach: as it provides an unbiased estimate for $ \Lambda $ \citep{Worsley1996c}.

\section{Further Discussion}

\subsection{Discussion of the computational aspects of convolution fields}\label{AA:convfields}
While one could think that the convolution approach would take up a lot of memory we were able to avoid this concern, by using optimization algorithms to find the peaks of the test-statistic. While this works well to test the global null hypothesis (or within a given region) it is a little more difficult to test the null within a given a voxel. Luckily this is more of a problem for very low sample sizes, where the test-statistic is quite rough, for which it is easier to generate high resolution lattices. For larger sample sizes only a slight resolution increase is required in order to avoid any conservativeness issues.


\section{Further Figures}\label{app:Figures}

\FloatBarrier
\subsection{Voxel-manifolds}
\FloatBarrier
\begin{figure}[h]
	\begin{center}
		\includegraphics[width=0.27\textwidth]{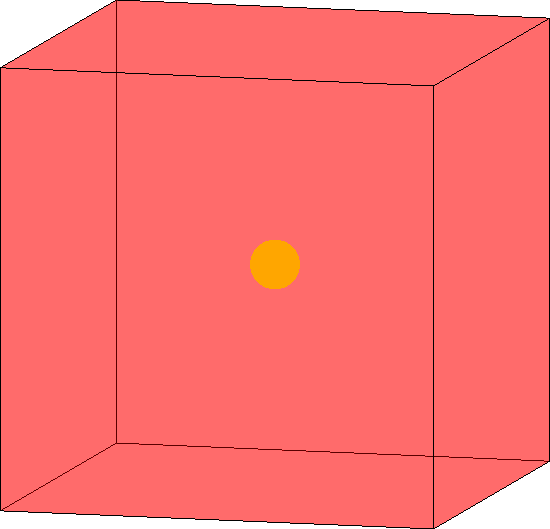}
		\quad
		\includegraphics[width=0.27\textwidth]{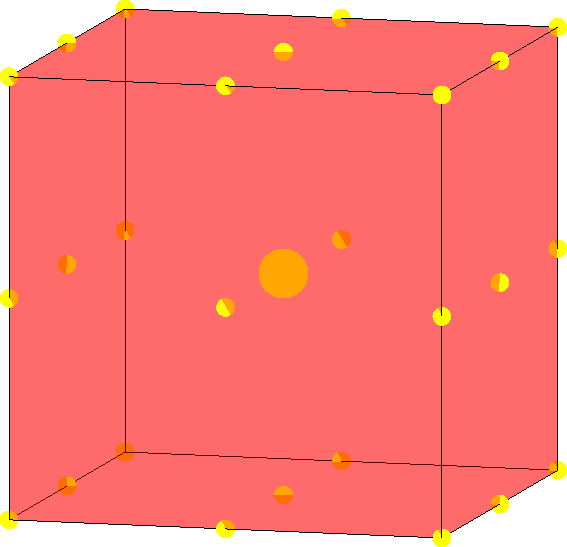}
		\quad
		\includegraphics[width=0.27\textwidth]{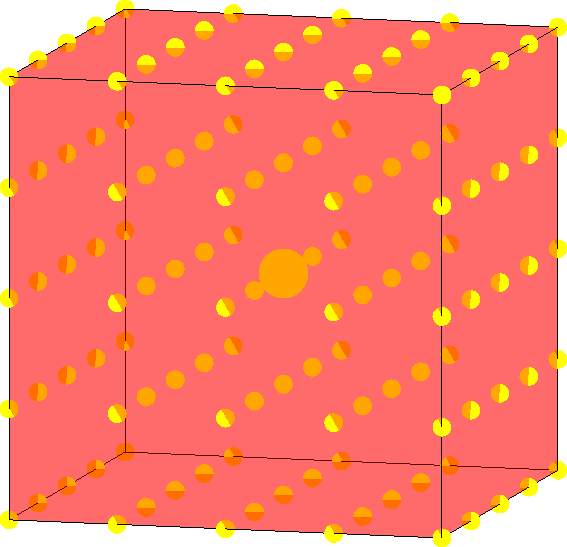}
	\end{center}
	\caption{Illustration of the added resolution for a 3D mask consisting of a single voxel. The set-up is the same as in Figure \ref{fig:resaddcomp}.}\label{fig:singlevox}
\end{figure}

\begin{figure}[h]
	\begin{center}
		\includegraphics[width=0.27\textwidth]{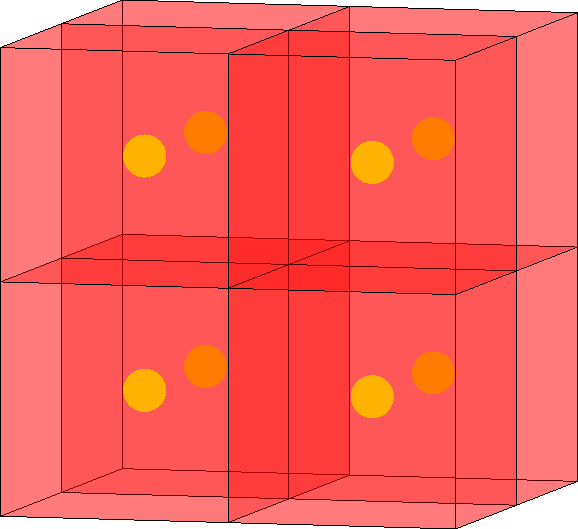}
		\quad
		\includegraphics[width=0.27\textwidth]{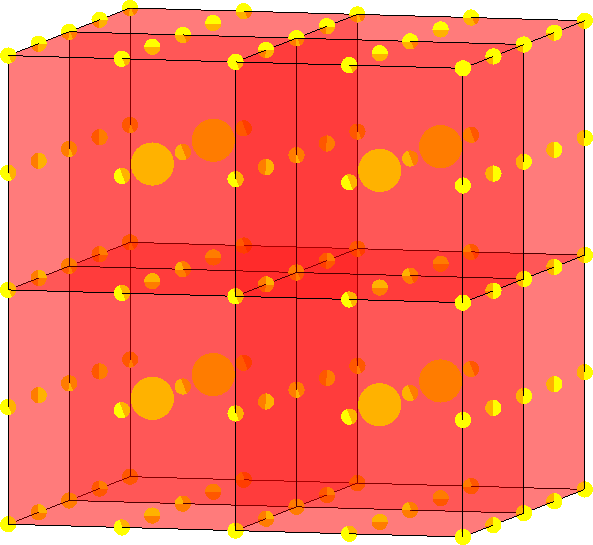}
		\quad
		\includegraphics[width=0.27\textwidth]{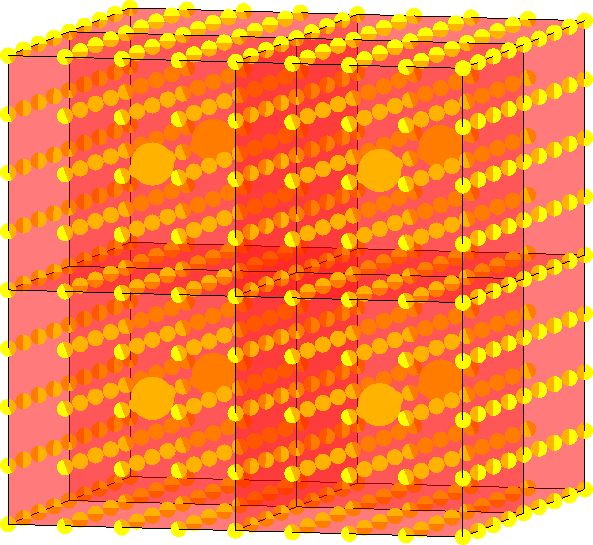}
	\end{center}
	\caption{Illustration of the added resolution for a 3D mask consisting of 8 voxels. The set-up is the same as in Figure \ref{fig:resaddcomp}.}\label{fig:2by2}
\end{figure}

\begin{figure}
		\begin{center}
		\includegraphics[width=0.49\textwidth]{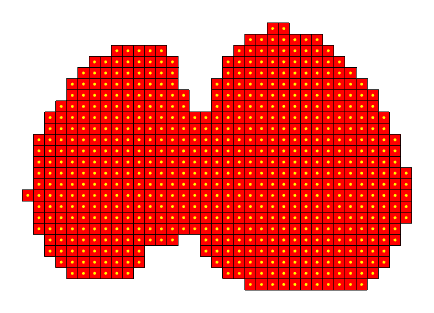}
		\includegraphics[width=0.49\textwidth]{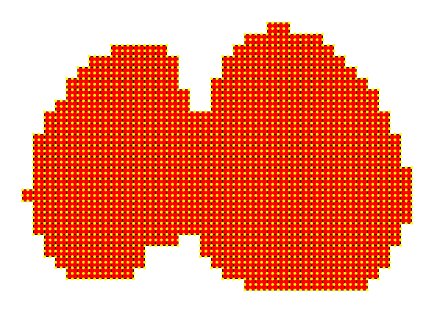}
	\end{center}
	\caption{Illustration of the added resolution for a 2D slice through the MNI mask. The red squares correspond to the voxel manifold and the yellow dots to the points of $ \mathcal{V}$ (left) and $ \mathcal{V}_1 $ (right).}\label{fig:2DMNI}
\end{figure}

\FloatBarrier
\subsection{Further LKC simulations}
\FloatBarrier
\begin{figure}[h!]
	\begin{center}
	\begin{subfigure}[b]{0.48\textwidth}
		\centering
		\includegraphics[width=\textwidth]{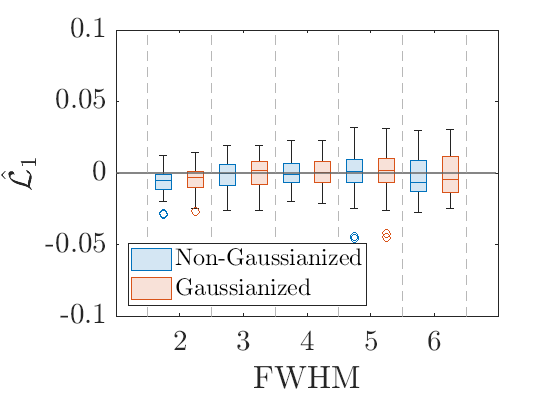}
	\end{subfigure}
	\begin{subfigure}[b]{0.48\textwidth}
		\centering
		\includegraphics[width=\textwidth]{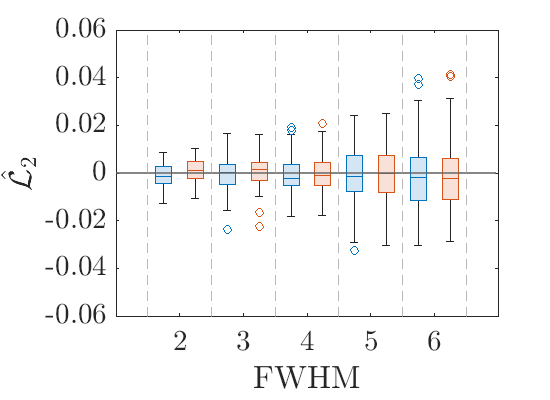}
	\end{subfigure}
\end{center}
	\caption{Dependence on the FWHM of the Gaussian smoothing kernel of the relative bias
			 of the LKC estimates using added resolution $1$
			 of the non-stationary	convolution field obtained
			 from the voxel data given by i.i.d. standard Gaussian random variables on
			 the MNI mask.
		The relative bias of the estimates of the LKCs without Gaussianization of the
		lattice data are shown in the blue boxes while the results with Gaussianization are
		the red boxes.}\label{fig:LKCplots_Gauss}
\end{figure}

\begin{figure}[h!]
	\begin{center}
	\begin{subfigure}[b]{0.48\textwidth}
		\centering
		\includegraphics[width=\textwidth]{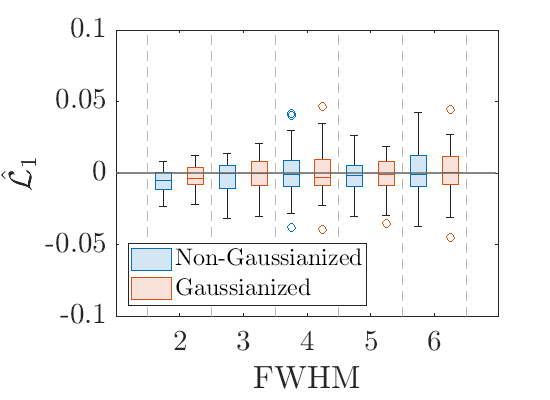}
	\end{subfigure}
	\begin{subfigure}[b]{0.48\textwidth}
		\centering
		\includegraphics[width=\textwidth]{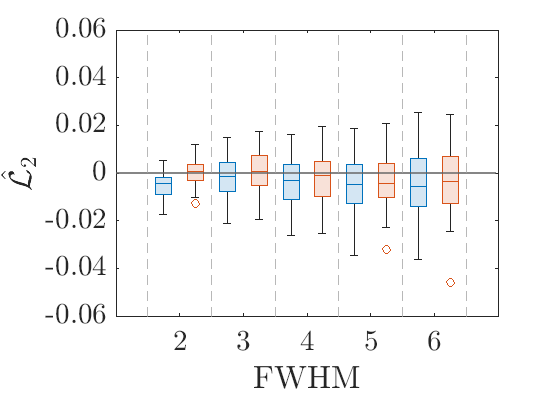}
	\end{subfigure}
\end{center}
	\caption{Dependence on the FWHM of the Gaussian smoothing kernel of the relative bias
			 of the LKC estimates using added resolution $1$
			 of the non-stationary	convolution field obtained
			 from the voxel data given by i.i.d. Laplace-distributed random variables
			 with zero-mean and scale parameter 3 on
			 the MNI mask.
		The relative bias of the estimates of the LKCs without Gaussianization of the
		lattice data are shown in the blue boxes while the results with Gaussianization are
		the red boxes.}\label{fig:LKCplots_laplace}
\end{figure}

\FloatBarrier
\subsection{Comparing EEC curves in the simulated data settings}
\FloatBarrier

\begin{figure}[h!]
	\begin{center}
		\begin{subfigure}[b]{0.3\textwidth}
			\centering
			\includegraphics[width=\textwidth]{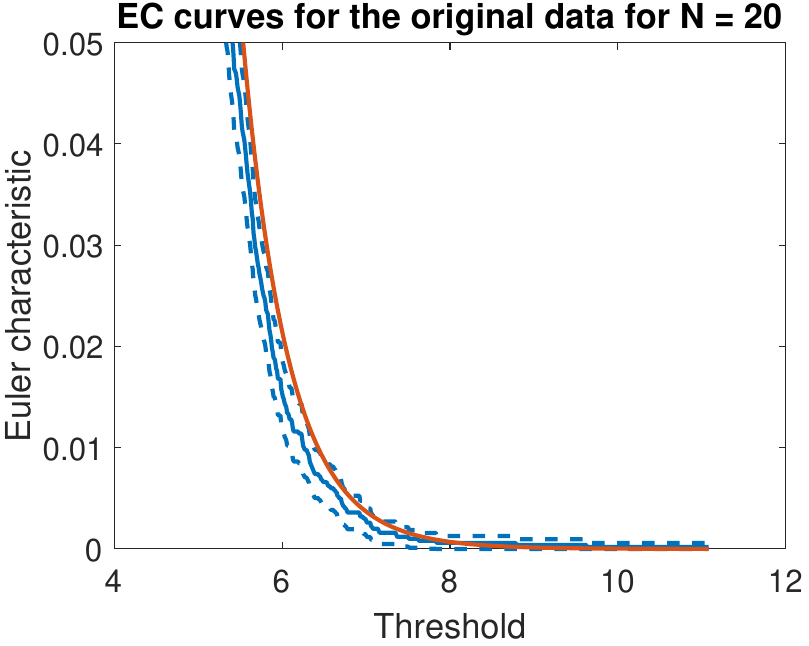}
		\end{subfigure}
		\begin{subfigure}[b]{0.3\textwidth}
			\centering
			\includegraphics[width=\textwidth]{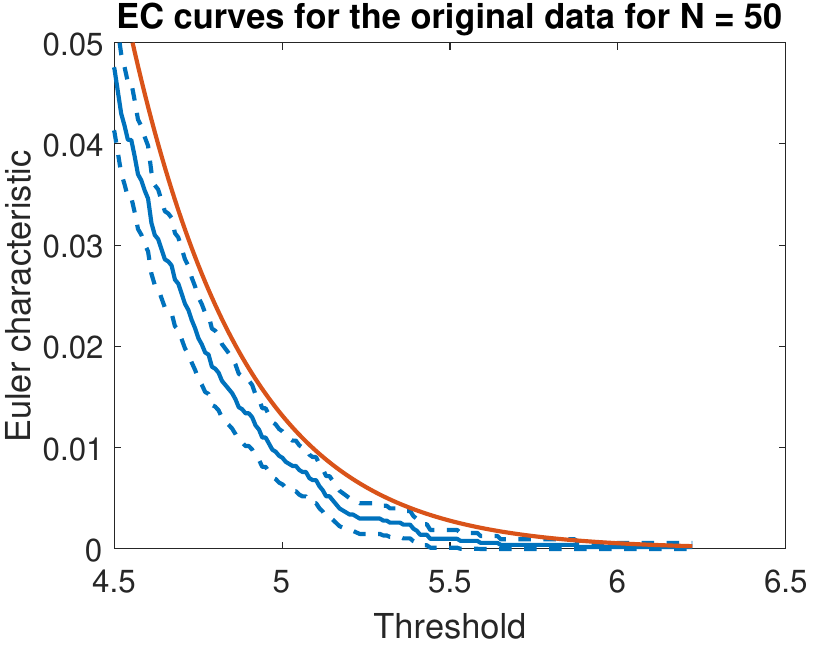}
		\end{subfigure}
		\begin{subfigure}[b]{0.3\textwidth}
			\centering
			\includegraphics[width=\textwidth]{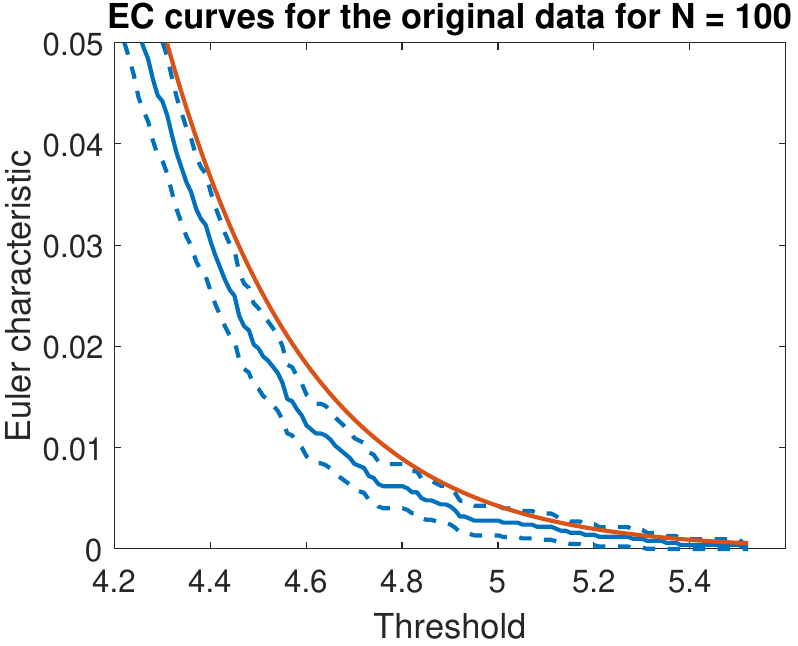}
		\end{subfigure}
	\end{center}
	\vskip\baselineskip
	\begin{center}
		\begin{subfigure}[b]{0.3\textwidth}
			\centering
			\includegraphics[width=\textwidth]{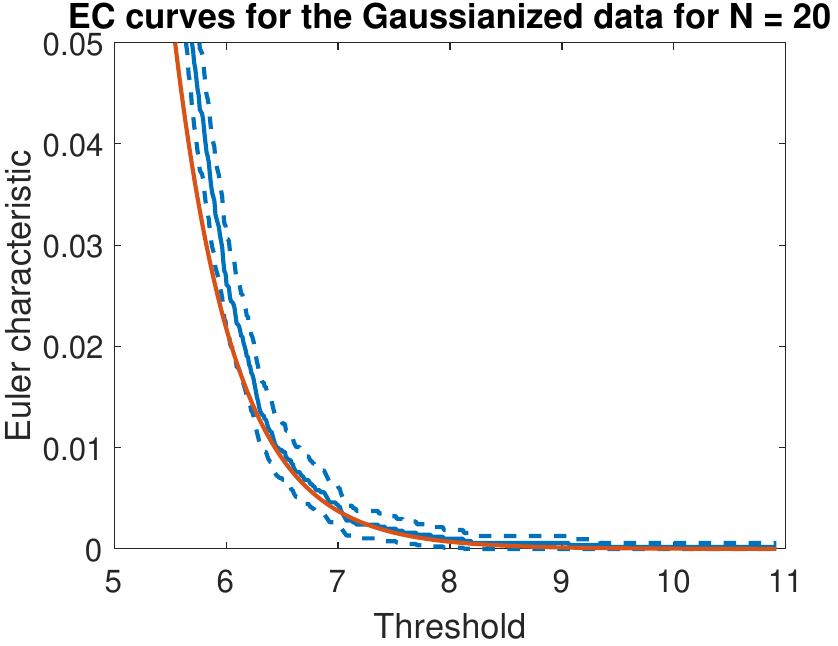}
		\end{subfigure}
		\begin{subfigure}[b]{0.3\textwidth}
			\centering
			\includegraphics[width=\textwidth]{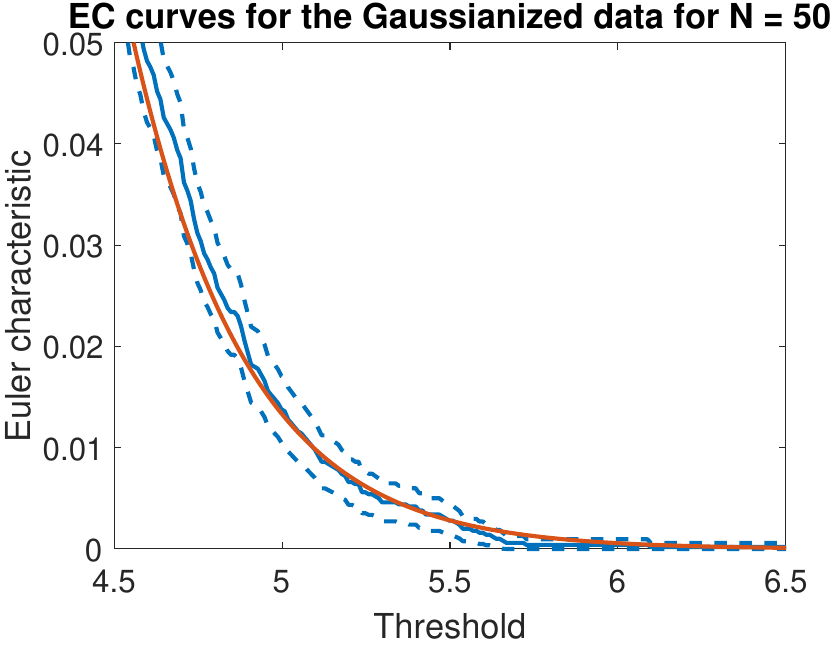}
		\end{subfigure}
		\begin{subfigure}[b]{0.3\textwidth}
			\centering
			\includegraphics[width=\textwidth]{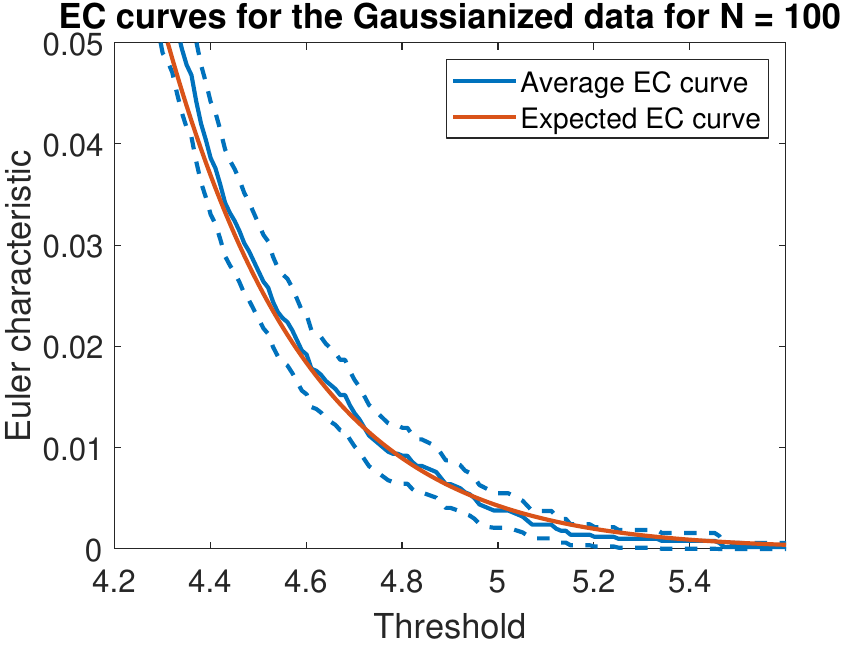}
		\end{subfigure}
	\end{center}
	\caption{Comparing the expected (shown in red) and empirical (shown in blue) tail EC curves for an applied smoothness of $ 4 $ FWHM per voxel. The empirical average curve is calculated as the average of the observed EC curves of the one-sample $ t $-statistics of the 5000 drawn samples of size $ N$. For the Gaussianized data (top row), given sufficiently many subjects, the empirical average EC curve is closely approximated by the EEC (calculated using the average of the LKCs over 5000 simulations) especially at high thresholds. For the original data (bottom row) the expected EC curve is not too far off but requires a much larger number of subjects for it to be an accurate approximation.}\label{fig:ECplots}
\end{figure}

\begin{figure}[h!]
	\begin{center}
		\begin{subfigure}[b]{0.3\textwidth}
			\centering
			\includegraphics[width=\textwidth]{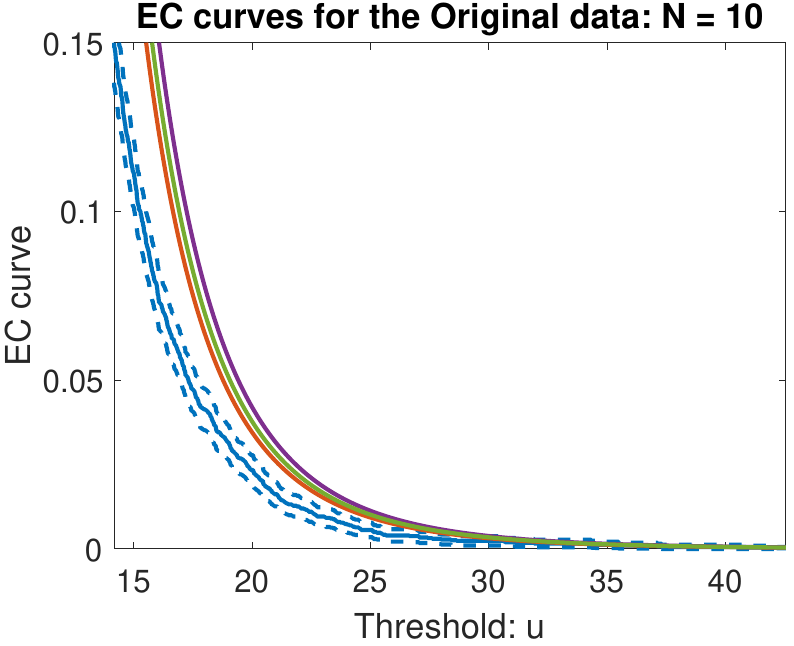}
		\end{subfigure}
		\begin{subfigure}[b]{0.3\textwidth}
			\centering
			\includegraphics[width=\textwidth]{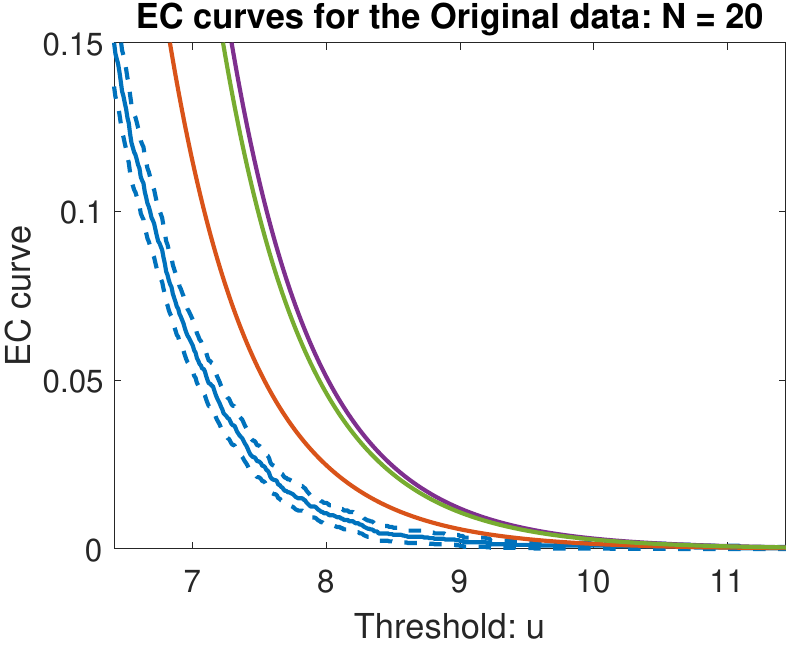}
		\end{subfigure}
		\begin{subfigure}[b]{0.3\textwidth}
			\centering
			\includegraphics[width=\textwidth]{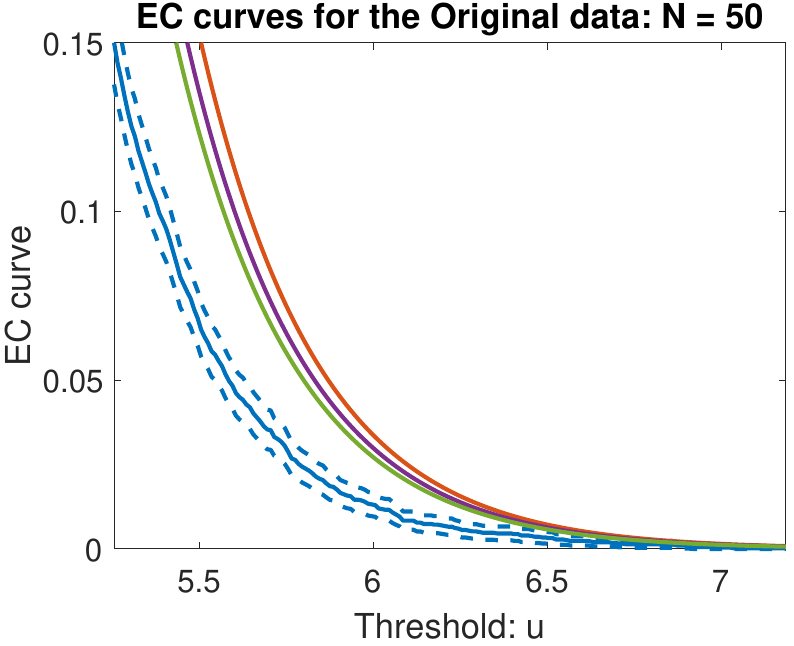}
		\end{subfigure}
	\end{center}
	\vskip\baselineskip
	\begin{center}
		\begin{subfigure}[b]{0.3\textwidth}
			\centering
			\includegraphics[width=\textwidth]{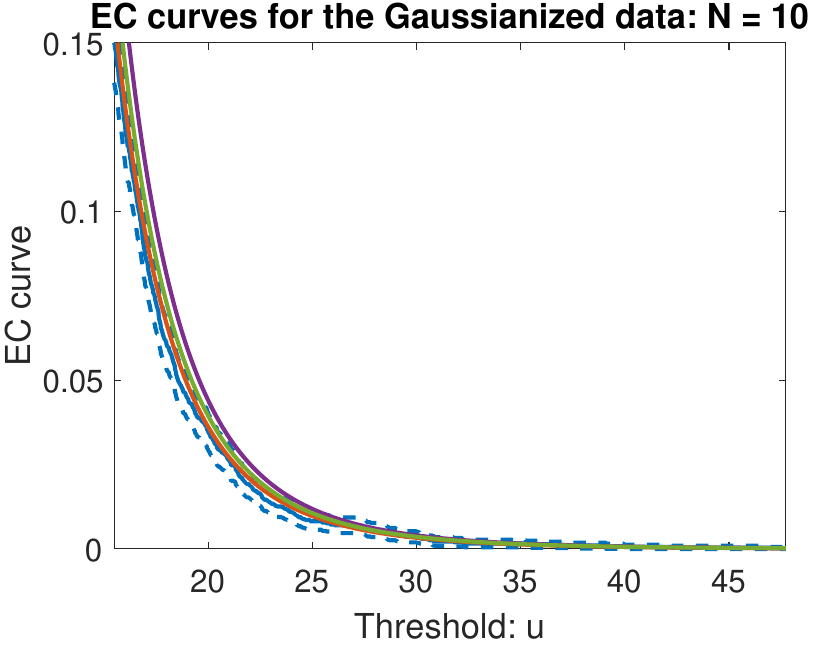}
		\end{subfigure}
		\begin{subfigure}[b]{0.3\textwidth}
			\centering
			\includegraphics[width=\textwidth]{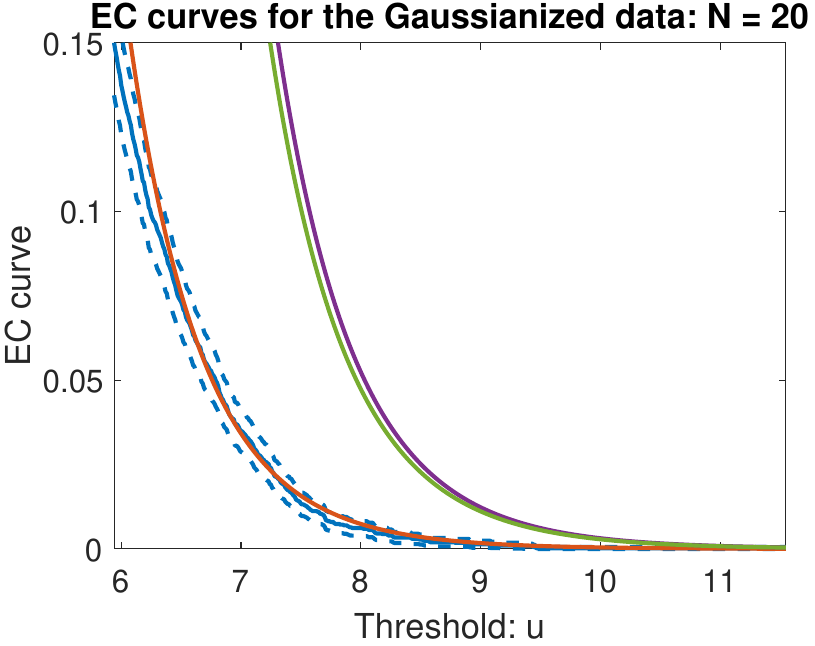}
		\end{subfigure}
		\begin{subfigure}[b]{0.3\textwidth}
			\centering
			\includegraphics[width=\textwidth]{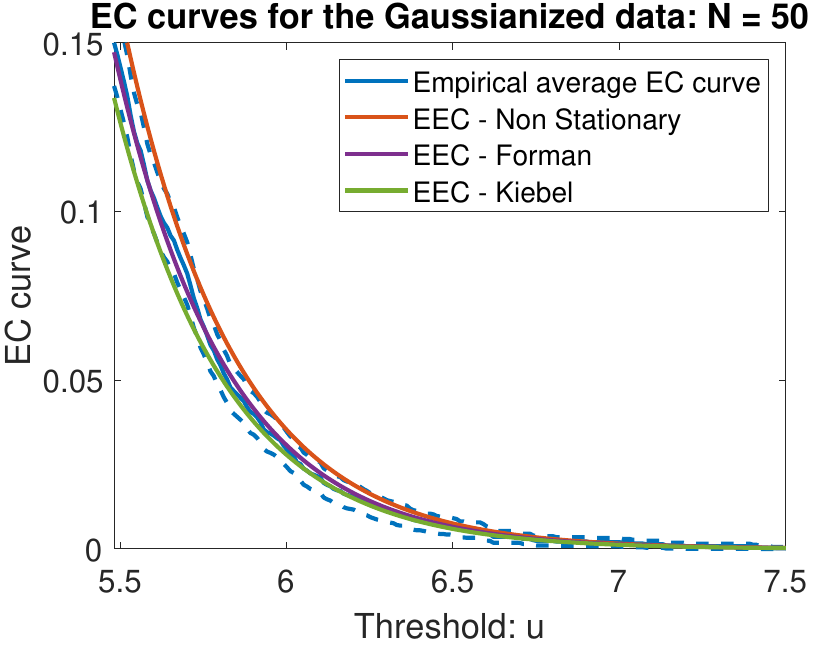}
		\end{subfigure}
	\end{center}	
	\caption{Comparing the expected and empirical tail EC curves in different settings using the resting state data for an applied smoothing level of 2 FWHM per voxel. The results are similar to Figure  \ref{fig:EECplotsFWHM5}.}\label{fig:EECplotsFWHM2}
\end{figure}

In Figure \ref{fig:ECplots} we plot the EEC curve (calculated using the average of the LKCs over the 5000 simulations as described in Section \ref{SS:Gauss}) and compare it to the empirical EC curve. For the original data the GKF breaks down at finite $N$, since the data is not Gaussian. This causes inaccuracies in EEC estimation, even at $ N = 100 $ and causes the EEC to be overestimated. This leads to conservative inference, as we shall see in the next section. Instead for the transformed data these curves closely match especially at high thresholds, which are the ones needed for FWER inference.

\FloatBarrier

